\documentclass{ametsocV6.1}
\usepackage{siunitx}

\title{Learned 1-D passive scalar advection to accelerate chemical transport modeling: a case study with GEOS-FP horizontal wind fields}

\authors{Manho Park,\aff{a} 
Zhonghua Zheng,\aff{b} 
Nicole Riemer,\aff{c} 
Christopher W. Tessum,\aff{a} \correspondingauthor{Christopher W. Tessum, ctessum@illinois.edu}
}

\affiliation{\aff{a}{Department of Civil and Environmental Engineering, University of Illinois Urbana-Champaign, Urbana, IL}\\
\aff{b}{Department of Earth and Environmental Sciences, The University of Manchester, Manchester M13 9PL, United Kingdom}\\
\aff{c}{Department of Atmospheric Sciences, University of Illinois Urbana-Champaign, Urbana, IL}\\
}

\abstract{We developed and applied a machine-learned discretization for one-dimensional (1-D) horizontal passive scalar advection, which is an operator component common to all chemical transport models (CTMs). 
Our learned advection scheme resembles a second-order accuracy, three-stencil numerical solver, but differs from a traditional solver in that coefficients for each equation term are output by a neural network rather than being theoretically-derived constants. 
We downsampled higher-resolution simulation results
---resulting in up to 16$\times$ larger grid size and 64$\times$ larger timestep—and trained our neural network-based scheme to match the downsampled integration data. 
In this way, we created an operator that is low-resolution (in time or space) but can reproduce the behavior of a high-resolution traditional solver. 
Our model shows high fidelity in reproducing its training dataset (a single 10-day 1-D simulation) and is similarly accurate in simulations with unseen initial conditions, wind fields, and grid spacing. 
In many cases, our learned solver is more accurate than a low-resolution version of the reference solver, but the low-resolution reference solver achieves greater computational speedup (500$\times$ acceleration) over the high-resolution simulation than the learned solver is able to (18$\times$ acceleration). 
Surprisingly,  our learned 1-D scheme---when combined with a splitting technique---can be used to predict 2-D advection, and is in some cases more stable and accurate than the low-resolution reference solver in 2-D. 
Overall, our results suggest that learned advection operators may offer a higher-accuracy method for accelerating CTM simulations as compared to simply running a traditional integrator at low resolution. 
} 

\begin{document}

\maketitle

%
%
%
\statement
	 Chemical transport modeling (CTM) is an essential tool to study air pollution. CTM simulation takes a long computing time. Modeling pollutant transport (advection) is the second most computationally intensive part of the model. Decreasing the resolution reduces the advection computing time, but decreases accuracy. We employed machine learning to reduce the resolution of advection while keeping the accuracy. We verified the robustness of our solver with several generalization testing scenarios. In our 2-D simulation, our solver showed a 100-time faster simulation with fair accuracy. Integrating our approach to existing CTM will enable much more simulations than now to study air pollution and to provide solutions.
%
%

%
\section{Introduction}

Atmospheric chemical transport models (CTMs) are a key research tool to study air pollution and to predict outcomes of pollution mitigation efforts. 
Currently available CTMs (e.g., GEOS-Chem \citep{bey2001global}, CMAQ \citep{us_epa_office_of_research_and_developmen_2022_7218076}, and WRF-Chem \citep{grell2005fully}) numerically represent physicochemical processes including emission, chemical reactions, transport, and deposition discretized across many grid boxes. 
This numerical discretization engenders a dilemma between accuracy and computational cost: higher-resolution simulations are more accurate but can be computationally intractable for some use cases. 
Parallel and distributed computation can reduce the overall execution time of simulations but add cost and complexity. 
For example, \citet{zhuang2020enabling} found that an $8\times$ increase in distributed computing power resulted in only a $4\times$ increase in speed for GEOS-Chem simulations.

Recent efforts have explored the use of machine-learned (ML) surrogate modeling to address computational costs in CTMs. 
Because the chemistry solver is the most expensive part of the atmospheric chemistry model, most ML research related to the acceleration of CTMs has focused on the chemistry operator \citep{keller2017machine, keller2019application,shen2022machine,kelp2020toward,kelp2022online,huang2022neural}. 
The advective transport operator is the second-most expensive module of the air quality model \citep{eastham2018geos}, but we are only aware of one existing study of machine-learned advection \citep{zhuang2021learned}. 
\citet{sturm2023advecting} accelerated the advection using by compressing the number of species, rather than accelerating the advection solver itself. 
However, recent progress in machine-learned computational fluid dynamics (CFD) and turbulence simulation \citep{kochkov2021machine, stachenfeld2021learned, brunton2020machine, vinuesa2022enhancing}, has laid a solid foundation for the acceleration of the transport operator in the context of air quality modeling.

In this study, we develop a machine-learned 1-D horizontal advection solver in the context of an air quality model. 
We build on the work of \citet{zhuang2021learned}, who trained a model using synthetic velocity fields within the rectangular computational domain with a fixed grid size. 
Here, as a step toward full-scale deployment in an ML advection operator in a CTM, we instead use real-world historical wind velocity data and grid cells that change size with changing latitude. 
Furthermore, we extensively explore the robustness of our approach by limiting the training data used and by testing its performance under a variety of different conditions. 
Altogether, the results herein characterize the promise of machine-learned advection operators for use in CTMs and outline the remaining challenges to their full-scale adoption.

The remaining parts of this paper are structured as follows: Section~\ref{sec:method} describes our methodology, including the numerical advection scheme used as a reference solver, the structure of the learned scheme which we train to match the results of our reference solver, and numerical experiments including generalization testing and 2-D application. 
Section~\ref{sec:result} presents the results of the experiments introduced in the methodology section. 
Section~\ref{sec:discussion} discusses strengths and limitations of our approach, and future research needs. 
Section~\ref{sec:conclusion} concludes this paper.

\section{Methodology}\label{sec:method}
\subsection{Numerical advection}
\subsubsection{Numerical scheme and dataset generation}\label{sec:numerical}

We implemented the so-called L94 advection scheme \citep{lin1994class} using the Julia scientific computing language \citep{bezanson2012julia}. 
The L94 advection scheme is a second-order-accurate van Leer-type advection scheme and has been used in a well-established chemical transport model, GEOS-Chem \citep{bey2001global}, when coupled with a nondirectional multidimensional splitting \citep{lin1996multidimensional}. 
An implementation in Julia—a differentiable programming language \citep{innes2019differentiable}—allows us to train machine-learned models by backpropagating error gradients  through multiple model time steps. 
Following \citet{lin1994class}, the equations for our 1-D numerical advection reference model are as follows:

Suppose we have a scalar field discretized across a 1-D grid, with concentration $\Phi_{i}^{n}$ in the i$^{\rm th}$ grid point at the $n^{\rm th}$ timestep. 
Then, the cell average value ($\Phi_{i+1/2}^{n}$) can be computed as Equation~\ref{eq:cellaverage} because the L94 scheme assumes the piecewise linear distribution of the scalar field inside a cell:
\begin{equation} \label{eq:cellaverage}
\Phi_{i+1/2}^{n} = (\Phi_{i}^{n} + \Phi_{i+1}^{n})/2.
\end{equation}
The ultimate purpose of this flux form operator is to calculate the next time cell average as in Equation~\ref{eq:nexttime}:
\begin{equation} \label{eq:nexttime}
\Phi_{i+1/2}^{n+1} = \Phi_{i+1/2}^{n} - \frac{\Delta t}{\Delta x} [f_{(i+1)} - f_{(i)}],
\end{equation}
where the flux ($f_{(i)}$) through the cell boundary can be expressed by Equation~\ref{eq:cellfluxpos} for $U_{i} \geq$ 0 or Equation~\ref{eq:cellfluxneg} for $U_{i} < 0$, where $U_{i}$ is the velocity at the i$^{\rm th}$ edge of the grid cell:
\begin{subequations}
\begin{eqnarray}
f_{(i)} = U_{i}[\Phi_{i-1/2}^{n} + \frac{\Delta \Phi_{i-1/2}^{n}}{2} (1 - C_{i}^{-})] \label{eq:cellfluxpos},\\
f_{(i)} = U_{i}[\Phi_{i+1/2}^{n} + \frac{\Delta \Phi_{i+1/2}^{n}}{2} (1 + C_{i}^{+})].  \label{eq:cellfluxneg}
\end{eqnarray}
\end{subequations}
Both $C_{i}^{-}$ and $C_{i}^{+}$ are Courant–Friedrichs–Lewy (CFL) numbers \citep{courant1928partiellen}, which can be calculated using Equation~\ref{eq:cflneg} for $U_{i} < 0$ and Equation~\ref{eq:cflpos} for $U_{i} \geq$ 0, respectively. 
Here $\Delta x_{i+-1/2}$ and $\Delta x_{i+1/2}$ are the grid spacing on the left and right sides of the $i^{\rm th}$ grid cell edge, and $\Delta t$ is the timestep. 
\begin{subequations}
\begin{eqnarray}
C_{i}^{-} = \frac{U_{i}\Delta t}{\Delta x_{i-1/2}}, \label{eq:cflneg}\\
C_{i}^{+} = \frac{U_{i}\Delta t}{\Delta x_{i+1/2}}. \label{eq:cflpos}
\end{eqnarray}
\end{subequations}
The key feature of this reference scheme is the method for deriving the derivative of the cell-averaged scalar, which is implemented using a monotonicity constraint (denoted ``mono5'' following \citet{lin1994class}) to ensure numerical stability as shown in Equation~\ref{eq:limiter}:
\begin{equation} \label{eq:limiter}
[\Delta \Phi_{i+1/2}]_\textnormal{mono5} = \textnormal{sign}([\Delta \Phi_{i+1/2}]_\textnormal{avg}) \times \textnormal{min}[|[\Delta \Phi_{i+1/2}]_\textnormal{avg}|, 2(\Phi_{i+1/2} - \Phi_{i+1/2}^\textnormal{min}), 2(\Phi_{i+1/2}^\textnormal{max} - \Phi_{i+1/2})].
\end{equation}
In Equation~\ref{eq:limiter}, $\Phi_{i+1/2}^\textnormal{max}$ and $\Phi_{i+1/2}^\textnormal{min}$ represent the local maximum and minimum, respectively, of $\Phi$ in the three-point stencil centered at $i+1/2$. 
Then, $[\Delta \Phi_{i+1/2}]_\textnormal{avg}$ is the average of the local spatial difference in $\Phi$ as in Equation~\ref{eq:meandelta}:
\begin{equation} \label{eq:meandelta}
[\Delta \Phi_{i+1/2}]_\textnormal{avg} = \frac{(\delta \Phi_{i} + \delta \Phi_{i+1})}{2},
\end{equation}
where
\begin{equation} \label{eq:delphi}
\delta \Phi_{i} = \Phi_{i+1/2}^{n} - \Phi_{i-1/2}^{n}.
\end{equation}

Our 1-D simulation domain is a straight line across the GEOS-FP $\SI{0.25}{\degree}$ $\times$ $\SI{0.3125}{\degree}$ North America nested grid \citep{geosfp} which covers $\SI{130}{\degree}$W--$\SI{60}{\degree}$W and $\SI{9.75}{\degree}$N--$\SI{60}{\degree}$N. 
We used the component of the ground-level wind field aligned with the domain line direction, using velocity data from $\SI{0.25}{\degree}$ $\times$ $\SI{0.3125}{\degree}$ GEOS-FP wind fields \citep{geosfp}. 
For example, if the line lay in the east-west direction, we would use the east-west component of the given wind field. 

\begin{figure}[htp]
    \centering
    \includegraphics[width=0.5\textwidth]{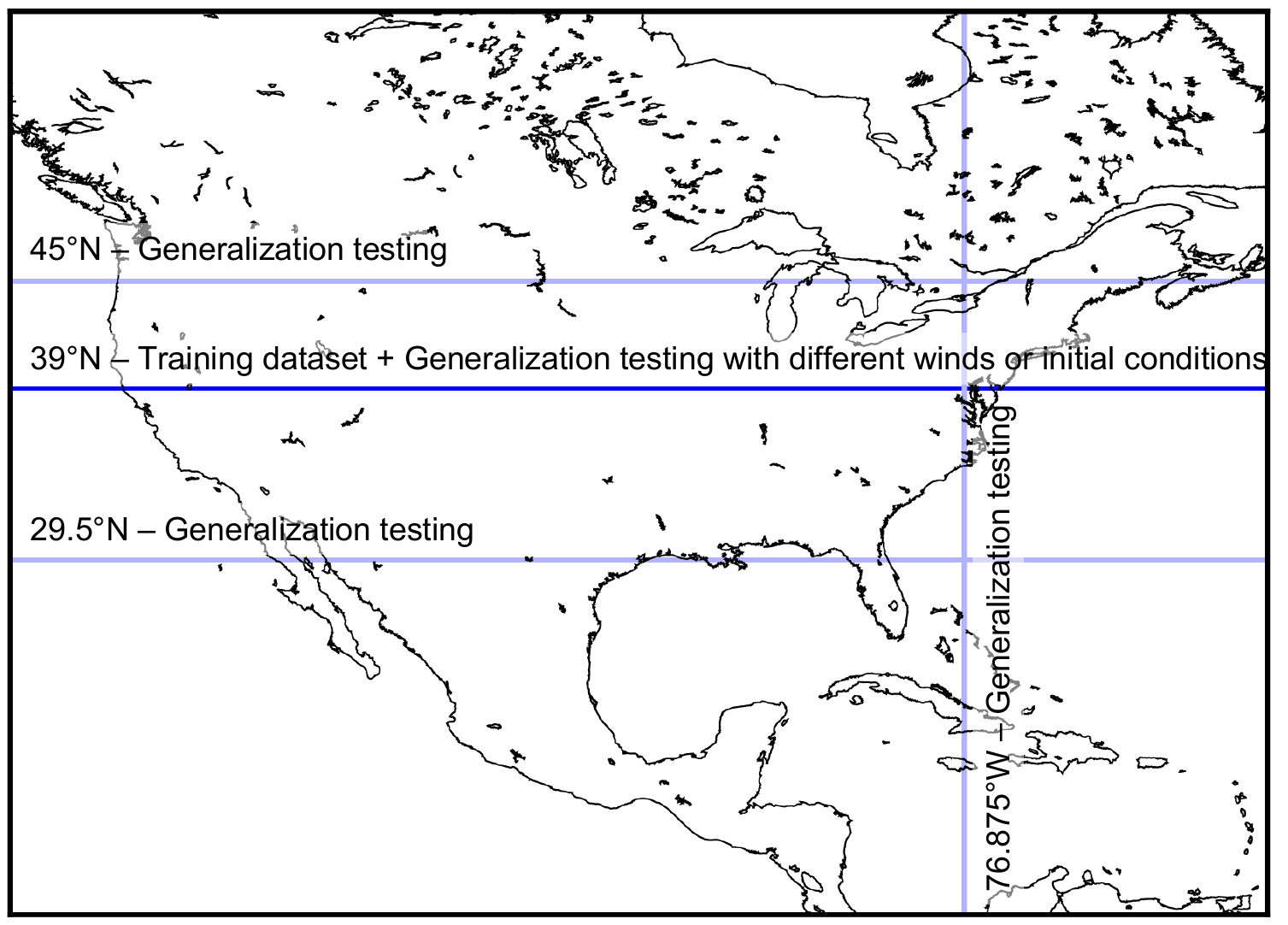}
    \caption{Visualization of the 1-D training and testing domain.}
    \label{fig:Map}
\end{figure}

We generated a training dataset by simulating advection through the East-West horizontal line on $\SI{39.00}{\degree}$N with $\SI{0.3125}{\degree}$ spatial resolution, using a 300-second timestep with the reference solver described above. 
The thick blue line in Figure~\ref{fig:Map} shows the training domain. 
The simulation period is the first 10 days of January 2019. 
We fed a square shape initial condition which has a 100~ppb mixing ratio on the central one-third of the domain and zero on the rest of the domain. 
We chose 100~ppb for the initial mixing ratio of the passive tracer as it is a typical mixing ratio for atmospheric pollutants and the default initial mixing ratio of the passive scalar in the GEOS-Chem transport tracer simulation. 
The square-shaped initial condition results in a challenging training dataset because the integrated wave will have a stiff gradient at the beginning of the simulation, which becomes smoother as the simulation progresses. 
We specified zero-gradient spatial boundary conditions. 
This results in simulations with 224 grid cells and 2880 integration timesteps. 

\subsubsection{Spatio-temporal coarsening}\label{sec:coarsening}

We down-sampled the scalar outputs of the advection simulation with factors of 1$\times$, 2$\times$, 4$\times$, 8$\times$ and 16$\times$ in space and  1$\times$, 2$\times$, 4$\times$, 8$\times$, 16$\times$, 32$\times$, and 64$\times$ in time. 
We applied both spatial and temporal down-sampling together, resulting in 35 different scenarios of data coarsening. 
We down-sampled the output of the reference simulation by averaging as shown in Figure~\ref{fig:Downsample}. 
At maximum coarsening in both space and time, the data dimensionality was reduced from 224~cells $\times$ 2,880~steps to 14~cells $\times$ 45~steps. 
We then trained machine-learning models to reproduce the down-sampled reference simulation.

\subsection{Development of the learned advection scheme}
\subsubsection{Model structure}\label{sec:modelstructure}
\begin{figure}[htp]
    \centering
    \includegraphics[width=\textwidth]{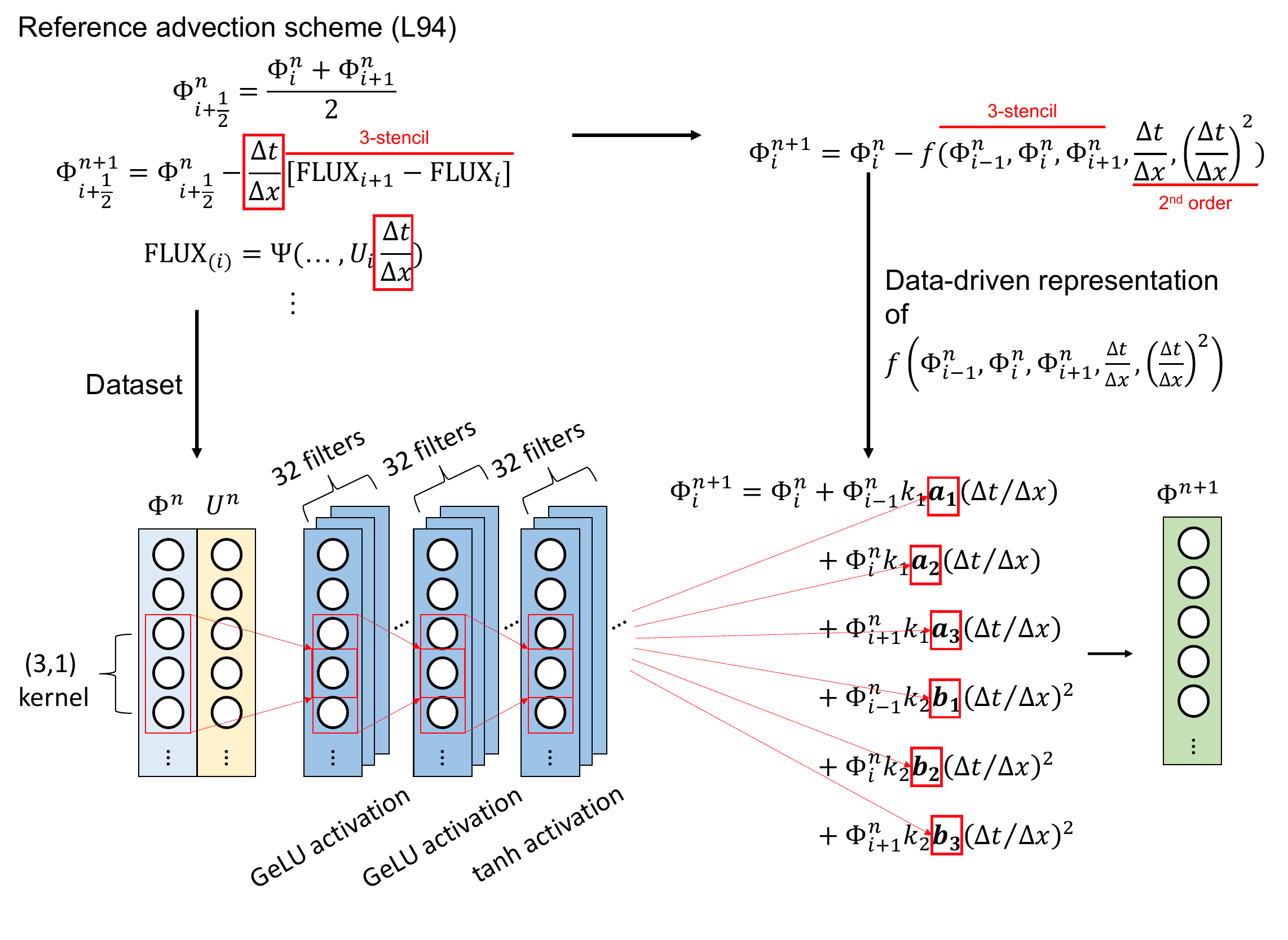}
    \caption{Design of the neural net-based models to emulate the L94 advection solver. $\Phi^{n}$ is the scalar field and $U^{n}$ is the velocity field.}
    \label{fig:ModelStructure}
\end{figure}

As illustrated in Figure~\ref{fig:ModelStructure}, we designed a physics-informed machine learning solver that is fed with scalar and velocity fields and returns the surrogate coefficients for numerical integration of a 3-stencil second-order-accurate scheme. 
We represented the temporal gradient using six terms describing 3-stencil first- and second-order derivatives as shown in Equation~\ref{eq:learnedsolver}:
\begin{equation} \label{eq:learnedsolver}
\Phi_{i}^{n+1} = \Phi_{i}^{n} + k_{1}\frac{\Delta t}{\Delta x} [a_{1}, a_{2}, a_{3}] \cdot [\Phi_{i-1}^{n}, \Phi_{i}^{n}, \Phi_{i+1}^{n}] + k_{2}\left(\frac{\Delta t}{\Delta x}\right)^{2} [b_{1}, b_{2}, b_{3}] \cdot [\Phi_{i-1}^{n}, \Phi_{i}^{n}, \Phi_{i+1}^{n}],
\end{equation}
where $k_{1}$ and $k_{2}$ are constant-gradient scaling factors (to help with training) and $a_i$ and $b_i$ are the surrogate coefficients generated by the learned solver. 
This formulation gives our learned scheme the flexibility to use different values of $\Delta$x, which can represent the variable relationship between grid cell size in meters and degrees at different latitudes. 
Overall, Equation~\ref{eq:learnedsolver} results in a machine-learned model with a strong inductive bias based on the mathematical framework described in Section~\ref{sec:numerical}. 

We used a 3-layer convolutional neural network (CNN) with 32~filters per layer to fit the surrogate coefficients. 
\citet{zhuang2021learned} used a 4-layer CNN with 32~filters per layer to emulate 1-D advection and a 10-layer CNN with 32~filters per layer for 2-D. 
We used 32~filters per layer as they did and reduced the number of layers from 4 to 3 to reduce computational cost. 
We used the Flux.jl \citep{innes2018fashionable} machine learning software library to implement the neural network. 
We trained a separate neural network for each combination of temporal and spatial coarsening. 
We used a default kernel size of 3, corresponding to a 3-stencil scheme, but for simulations where the maximum CFL number is larger than 1, we increased the CNN kernel size (and stencil size) to 5, 9, 17, or 31 to provide a suitable spatiotemporal information horizon. 
For example, if the maximum CFL number of the downsampled velocity field is between 1 and 2, we set the kernel size to be five (i.e., two cells on the left side, one in the center, and two~cells on the right side) to allow our algorithm to access information about all scalar values which may pass through the grid cell of interest during a given timestep. 
To facilitate training our neural network across the stiff gradients that characterize this system, we employed the Gaussian error linear unit (GeLU; \citet{hendrycks2016gaussian}) activation function as suggested by \citet{kim2021stiff}. 
For the final activation layer, we used hyperbolic tangent to limit the rate of potential error growth. 
To normalize the temporal gradient for each term, we multiplied scaling factors on each term ($k_1$ and $k_2$ in Equation~\ref{eq:learnedsolver}), with magnitudes similar to the inverse of $\Delta t/\Delta x$ and $(\Delta t/\Delta x)^2$. 
In this study, we used 100 for $k_1$, and 10,000 for $k_2$, because $\Delta t = 300$~seconds and $\Delta x \approx 30,000$~km in the baseline resolution. 

\subsubsection{Neural net training}\label{sec:training}
We trained a separate neural network for each coarse-graining resolution listed in Section~\ref{sec:coarsening}. 
We trained our neural network to minimize the absolute error of a 10-step prediction following \citet{zhuang2021learned}'s suggestion and to reflect the temporal variation of the advected scalar field. 
(This means that each gradient descent step involves running the model forward for 10~steps with the current neural network weights, calculating the gradient of the average error among the 10 steps with respect to the weights, and then updating the weights in the opposite direction of the gradient.) 
In our experiments, the use of a single-timestep loss function resulted in instability or mode collapse. 
Input feature standardization was performed by multiplying the scalar array by $10^7$ and dividing the velocity array by 15 to ensure both scalar and velocity would be bound to a scale of $10^0$ for compatibility with the default parameters of the optimizer that we use. 
We introduced uniformly-distributed noise with a maximum value of $3\times10^{-4}$~times the magnitude of the initial condition into the scalar array to let the model learn the physical patterns behind the noise and prevent error accumulation, as shown by \citet{stachenfeld2021learned}. 
We used the Adam optimizer \citep{kingma2014adam} with the default parameters in Flux.jl. 
We modified the learning rate ($\rho$) to decay with the number of training epochs as in Equation~\ref{eq:decay}:
\begin{equation}\label{eq:decay}
\rho = \frac{2 \times 10^{-3}}{1+\rm epoch}.
\end{equation}
The purpose of the decaying learning rate is to encourage model convergence. 
We trained the model for 100 epochs by default and saved the model parameters every epoch. 
When the local minimum of training loss was observed within 100 epochs, we chose the training epoch as the best epoch. 
When the training loss was still decreasing after 100 epochs, we continued training until the loss stopped decreasing. 
The training epoch with the best performance for each resolution is summarized in Table~\ref{table:TrainEpoch}.

After training the model, we evaluated the correctness of our learned schemes in emulating the reference solver. 
To do this, we fed our learned models with the same initial conditions and integrated them with the same velocity fields as were used for the reference simulation. 
We evaluated the performance of the learned scheme using three different statistics: mean absolute error, root mean square error, and R$^{2}$, all averaged throughout the entire simulation rather than for a single timestep. 
Since error tends to accumulate with time integration, the scalar fields that the scheme faces in each step are never exactly the same as the scalar fields fed in the training procedure. 
Therefore, even though we use the same initial conditions and velocity fields used in the training dataset, the recurrent input data in each model simulation is somewhat distinct from the reference simulation data, so we are not testing on the training data here, but further generalization tests are described below. 

We evaluated the computational acceleration by the learned solver against the finest reference solver. 
The computational time to integrate the same duration (10 days) by each solver was measured using a single CPU (dual 6248 Cascade Lake CPU) within an HPE Apollo 6500 node installed in the University of Illinois Campus Cluster. 
We used the $@benchmark$ macro of BenchmarkTools.jl \citep{benchmarktools} to collect the statistics in time evaluation. 
We calculated the speedup as shown in Equation~\ref{eq:speedup}:

\begin{equation} \label{eq:speedup}
{\rm speed up} = \frac{{\rm mean}({\rm time_{learned}})}{{\rm mean}({\rm time_{baseline}})},
\end{equation}
where ${\rm time_{learned}}$ is the computational time spent by the learned solver and ${\rm time_{baseline}}$ is the computational time spent by the reference solver in the finest resolution. 
When the learned solver is faster than the baseline, the speedup is larger than 1.

\subsection{Generalization ability testing}\label{sec:generalization}
We evaluated the generalization ability of learned schemes against a longer time span (3$\times$ longer than the training dataset), different initial conditions (Dirac-delta shape and Gaussian distribution shape), and different wind fields. 
The tests on the different wind fields can be further categorized by seasons (the first 10~days in April, July, and October to represent Spring, Summer, and Fall respectively), latitudes ($\SI{29.5}{\degree}$N and $\SI{45}{\degree}$N), and domain direction (longitudinal line at $\SI{76.875}{\degree}$W using the north-south wind velocity component). 
The pale blue lines in Figure~\ref{fig:Map} show the spatial domains of the generalization tests. 
For the longer-term stability test, we simulated scalar advection from January 1st to 30th. 
For other tests, the integration time span is 10~days as in the training data.

Each test has a different purpose. 
The longer-duration simulations are used to evaluate the degradation of predictive performance over longer time horizons. 
Changing the shape of the initial conditions tests whether the learned schemes were over-fitted to the initial condition shape used in the training dataset. 
The Dirac-delta initial condition test investigates the performance of the learned scheme with extremely large spatial gradients. 
The purpose of using different velocity fields is to assess the robustness of the learned scheme on the different wind patterns. 
Finally, the tests with changes in latitude and domain direction evaluate the learned schemes' ability to work with different grid sizes. 
For context, the grid size of the training set ($\Delta$x, $\SI{0.3125}{\degree}$ in $\SI{39}{\degree}$N) is approximately 27.0~km, while $\Delta$x in $\SI{29.5}{\degree}$N and $\SI{45}{\degree}$N are 24.6 and 30.3~km, respectively. 
The longitudinal grid size, $\Delta$y—$\SI{0.25}{\degree}$—is approximately 27.8~km at all longitudes.

\subsection{Multidimensional splitting}\label{sec:multidimensional}
We implemented 2-D advection in the same way for both the reference solver and the learned scheme developed in this study. 
In both cases, 1-D advection is converted to 2-D using non-directional splitting as described in \citet{lin1996multidimensional} and reproduced in Equation~\ref{eq:splitting}:
\begin{equation}\label{eq:splitting}
Q^{n+1} = Q^{n} + F(Q^{n}+\frac{1}{2}G(Q^{n})) + G(Q^{n}+\frac{1}{2}F(Q^{n}))
\end{equation}
where $Q^n$ is the scalar value in a specific point on the 2-D domain at time $n$, and $F$ and $G$ are the x-axis and y-axis discretization operators, respectively. 
This means that for the learned solver, we used models to make 2-D predictions that were only trained to make 1-D predictions (at the same spatial and temporal resolution).
All simulations were performed over the GEOS-FP $\SI{0.25}{\degree}$ $\times$ $\SI{0.3125}{\degree}$ North America nested grid ($\SI{130}{\degree}$W--$\SI{60}{\degree}$W, $\SI{9.75}{\degree}$N--$\SI{60}{\degree}$N). 
The simulation period is the first 10~days in January 2019. 
The baseline resolution is $\SI{0.25}{\degree}$ $\times$ $\SI{0.3125}{\degree}$ in space, and 5 minutes in time. 
Like the 1-D study, coarse-graining was performed by doubling the spatial grid size and the timestep several times from $\SI{0.25}{\degree}$ $\times$ $\SI{0.3125}{\degree}$ $\times$ 5~minutes and up to $\SI{4.0}{\degree}$ $\times$ $\SI{5.0}{\degree}$ $\times$ 5~hours 20~minutes.

\section{Results}\label{sec:result}
\subsection{Model performance in emulating the training dataset}\label{sec:trainresult}

\begin{figure}[htp]
    \centering
    \includegraphics[width=\textwidth]{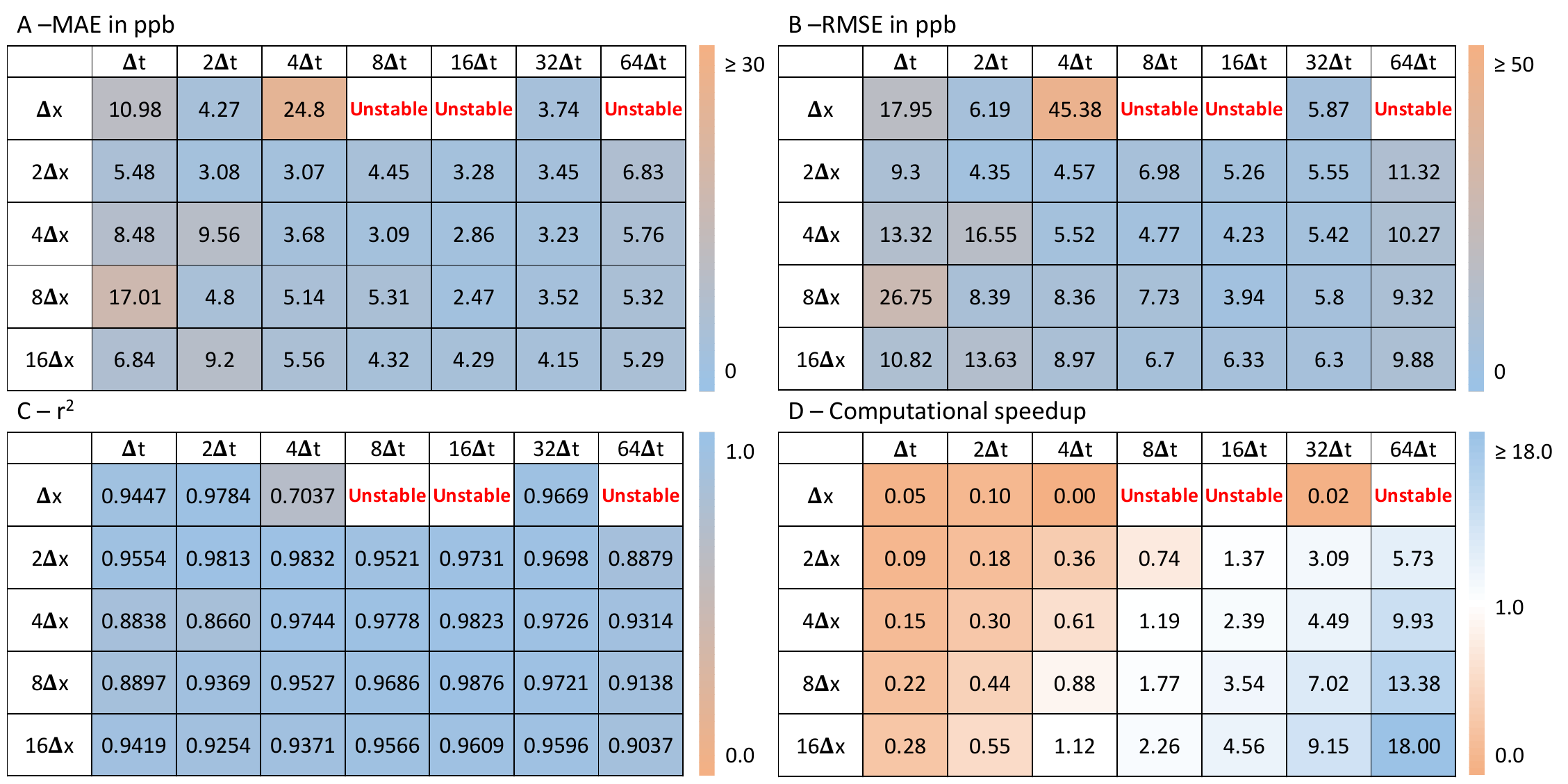}
    \caption{Performance of the \textbf{learned solver} in emulating the training dataset at different coarse-graining resolutions as compared to the highest-resolution reference solver simulation. Conditions where integration failed are marked ``Unstable''.}
    \label{fig:TrainStats}
\end{figure}

Figure~\ref{fig:TrainStats} summarizes trained model performance in integrating the training dataset with different levels of spatial and temporal coarsening. 
There are three cases where the learned solver returned unstable outputs, which are discussed in the final paragraph of this section. 
Except for these unstable cases, the overall errors are fairly small. 
MAEs and RMSEs range from 2.47
--24.8~ppb and 3.94--45.38~ppb, respectively (Figure~\ref{fig:TrainStats}\textbf{A} and \textbf{B}). 
MAE and RMSE values are mostly less than 10~ppb. $R^{2}$ values are mostly higher than 0.87 with a worst case of 0.70 (Figure~\ref{fig:TrainStats}C). 
The maximum computational speedup compared to the high-resolution reference model is 18.0$\times$ at the coarsest spatial and temporal resolution. 
However, under many conditions, our ML model is slower than the reference model owing to the computational intensity of the neural network. 
We find that temporal coarsening reduces computation time proportionately to the coarsening factor (e.g., 4$\Delta t$ is $2\times$ faster than 2$\Delta t$). 
Spatial coarsening also reduces computation time, but sub-proportionally to the coarsening factor (e.g., 4$\Delta x$ is $<2\times$ faster than 2$\Delta x$) because operations that are sequential in space can leverage the single instruction multiple data (SIMD) compiler optimizations whereas operations that are sequential in time cannot. 
The statistics of the computational time by the learned solvers and the reference solvers are given in Tables~\ref{table:time_learned} and \ref{table:time_reference}.

Computational speedup values reported in Figure~\ref{fig:TrainStats} and elsewhere are computed against our Julia implementation of the reference model described in Section~\ref{sec:numerical}. 
As a sense-check to demonstrate that any ML model speedups that we observed are not owing to an overly slow implementation of the reference model, we compare the execution time of our reference model to that of a GEOS-Chem ``transport tracer'' configuration, which only computes advection. We turned off other modules including chemistry, depositions, radiation, convection, boundary layer mixing, etc. 
On a single CPU in the Illinois Campus Cluster hardware described above, the GEOS-Chem ``transport tracer'' configuration in the nested $\SI{0.25}{\degree}$ $\times$ $\SI{0.3125}{\degree}$ North America domain with $\SI{4.0}{\degree}$ $\times$ $\SI{5.0}{\degree}$ global simulation for providing boundary conditions takes 53,677~seconds whereas the global $\SI{4.0}{\degree}$ $\times$ $\SI{5.0}{\degree}$ simulation takes 613~seconds for 10~days simulation. 
Therefore, the computational time for a 10-day simulation in nested North America is to be 53,065~seconds, or 1.89~microseconds per grid cell per timestep using a single CPU core, whereas our reference solver takes 0.06~microseconds per grid cell per timestep using a single CPU core. 
Although this comparison does not show that our reference solver is faster than GEOS-Chem—a ``transport tracer'' simulation does in fact not only do advection so it is not a fair comparison—but it does suggest that our reference algorithm is not unreasonably slow.
 
\begin{figure}[htp]
    \centering
    \includegraphics[width=\textwidth]{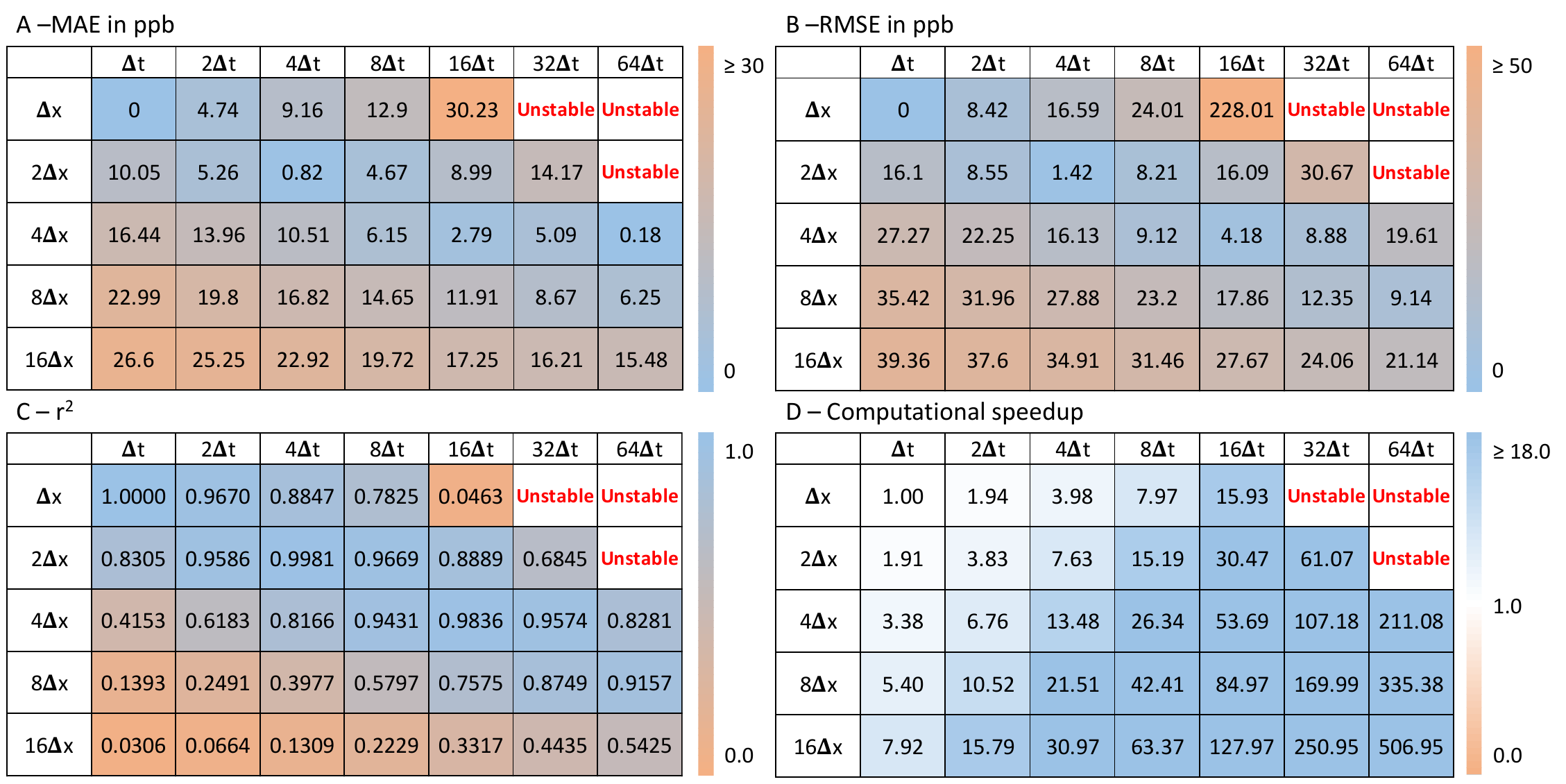}
    \caption{Performance of the \textbf{reference solver} in coarse-graining the fine resolution training dataset in different resolutions as compared to the highest-resolution reference solver simulation. Conditions where integration failed are marked ``Unstable''.}
    \label{fig:NumericalStats}
\end{figure}

To evaluate the benefit of our learned advection model, we compare its performance with simulations using the reference method run at spatial and temporal resolutions matching those of the learned models. 
Figure~\ref{fig:NumericalStats} summarizes the performance of the lower-resolution reference solver. 
Interestingly, the coarsening with the reference solver appears to have a preferential regime that lies on the diagonal elements in the performance matrix (e.g., 2$\Delta x$-2$\Delta t$, 4$\Delta x$-4$\Delta t$, etc). 
Previous research has suggested that shorter timesteps unconditionally result in higher model accuracy \citep{philip2016sensitivity}, but results here may provide contradictory evidence. 
Our results instead suggest that matching temporal and spatial coarsening may in some cases enhance the performance of the model as well as the computational speed, possibly because temporal coarsening averages out high-frequency features, which could trigger numerical diffusion in low spatial resolution. 
Also, the native timestep in the GEOS-FP wind field is 60 minutes, thus, temporal coarsening from 5~minutes to a longer step can be beneficial by reducing unnecessary computation and minimizing possible numerical error accumulation. 
Regardless, additional investigation is required to fully understand the phenomenon we observe here. 

When both solvers are run at the same resolution, the computational speedup by the reference solver is higher than that by the learned solver, owing to the computational complexity of the neural network (Figures~\ref{fig:TrainStats}D and \ref{fig:NumericalStats}D). 
For resolutions where we use a 3-layer CNN, the low-resolution reference solver was 18.7--28.5~times faster than the learned solver. 
For the (1$\Delta x$, 4$\Delta t$) and (1$\Delta x$, 32$\Delta t$) resolutions, we add an additional CNN layer to prevent instability, resulting in learned solvers that are $\sim$1,000$\times$ slower than the low-resolution reference solvers.
Given the fact that the reference solver with certain coarsening factors such as (8$\Delta x$, 64$\Delta t$) could achieve much faster speedup than the maximal speedup from the learned solver as well as favorable accuracy, we cannot say that the learned solver in this study is strictly superior to the traditional solver. 
Instead, we would like to argue that the learned solver presented here could be a complementary option in cases where the traditional solver does not work well. 
For example, the reference solver does not work well in cases with large grid cells combined with short timesteps (Figure~\ref{fig:NumericalStats})---such as nested simulations with coarse global domains but small timesteps to match the higher-resolution inner domains---but the learned solver can still operate with high accuracy in this regime. 
GEOS-Chem Classic operates the transport with 10 min by default in the global simulation and 5 min by default in the nested simulation. 
Using our learned solver, we can increase the time interval to even more than 10 minutes. 

\begin{figure}[htp]
    \centering
    \includegraphics[width=\textwidth]{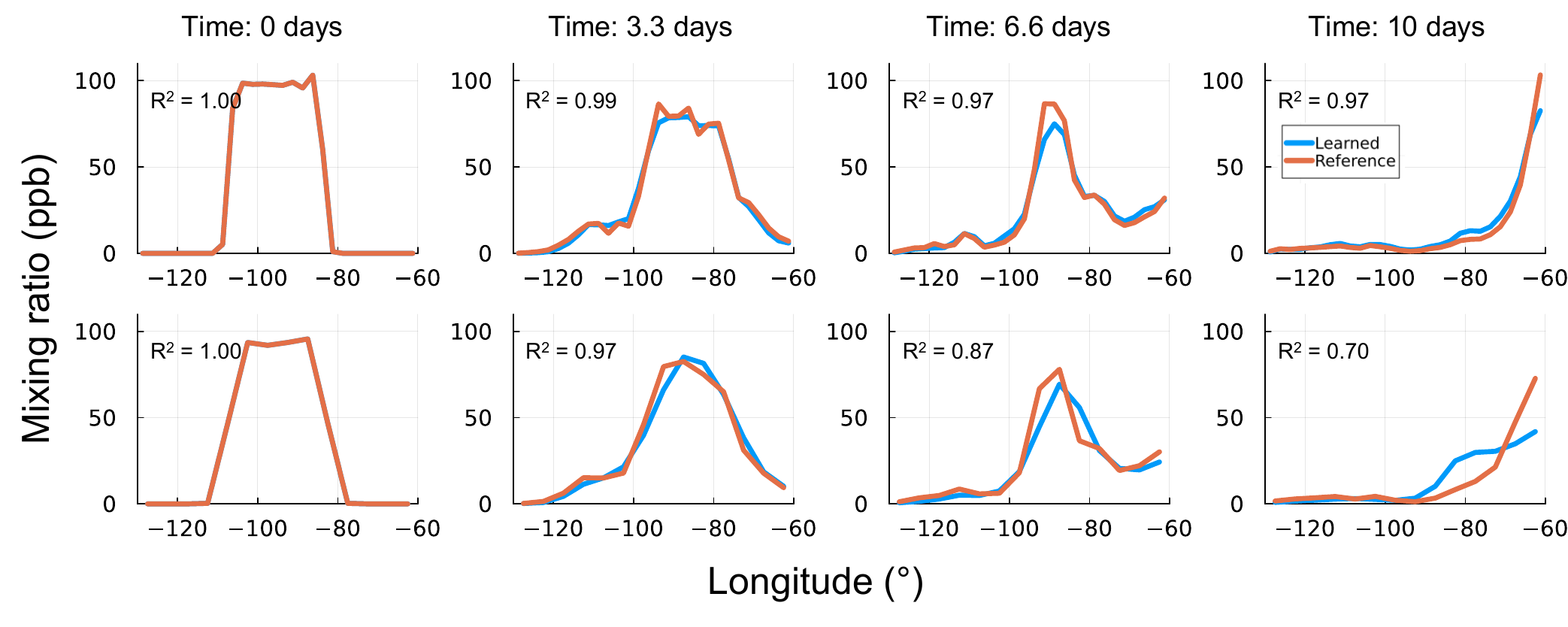}
    \caption{Time series representation of the learned emulation of the training dataset. The top row is 8$\Delta x$, 16$\Delta t$ and the bottom row is 16$\Delta x$, 64$\Delta t$. The initial shapes are slightly different from the square initial condition because of coarse-graining.}
    \label{fig:TimeSeriesTrain}
\end{figure}

The learned solvers with the highest accuracy predicting the training dataset (8$\Delta x$, 16$\Delta t$) and largest acceleration (16$\Delta x$, 64$\Delta t$) are shown in Figure~\ref{fig:TimeSeriesTrain}, which shows that the spatial patterns with time evolution are well-described by the learned schemes in both resolutions. 
Accuracy is not perfect, and deviations are more noticeable in (16$\Delta x$, 64$\Delta t$), but $R^2$ values combined across all timesteps are high in both cases (0.99 and 0.90, respectively). 

We are unable to successfully train the model with the coarsening factors of (1$\Delta x$, 8$\Delta t$), (1$\Delta x$, 16$\Delta t$), and (1$\Delta x$, 64$\Delta t$). In these cases, the learned model simulations become numerically unstable. 
One possible explanation is that those regimes have large CFL numbers due to large temporal coarsening factor relative to that in space. 
(The low-resolution reference model also become unstable under similar conditions.) 
The maximum CFL numbers in (1$\Delta x$, 8$\Delta t$), (1$\Delta x$, 16$\Delta t$), and (1$\Delta x$, 64$\Delta t$) are 1.82, 3.63, 7.26, and 14.52, respectively. 
The maximum CFL numbers in all the tested coarsening cases are summarized in Table~\ref{table:MaxCFL}. 
We use larger stencils and larger neural networks in such cases to account for the increased spatio-temporal transport horizon; this solved the problem for (1$\Delta x$, 4$\Delta t$) and (1$\Delta x$, 32$\Delta t$) but not for the cases above. 
Resolving this instability within a computationally efficient framework is a topic for future research.

\subsection{Generalization ability}\label{sec:generalizationresult}
\begin{figure}[htp]
    \centering
    \includegraphics[width=\textwidth]{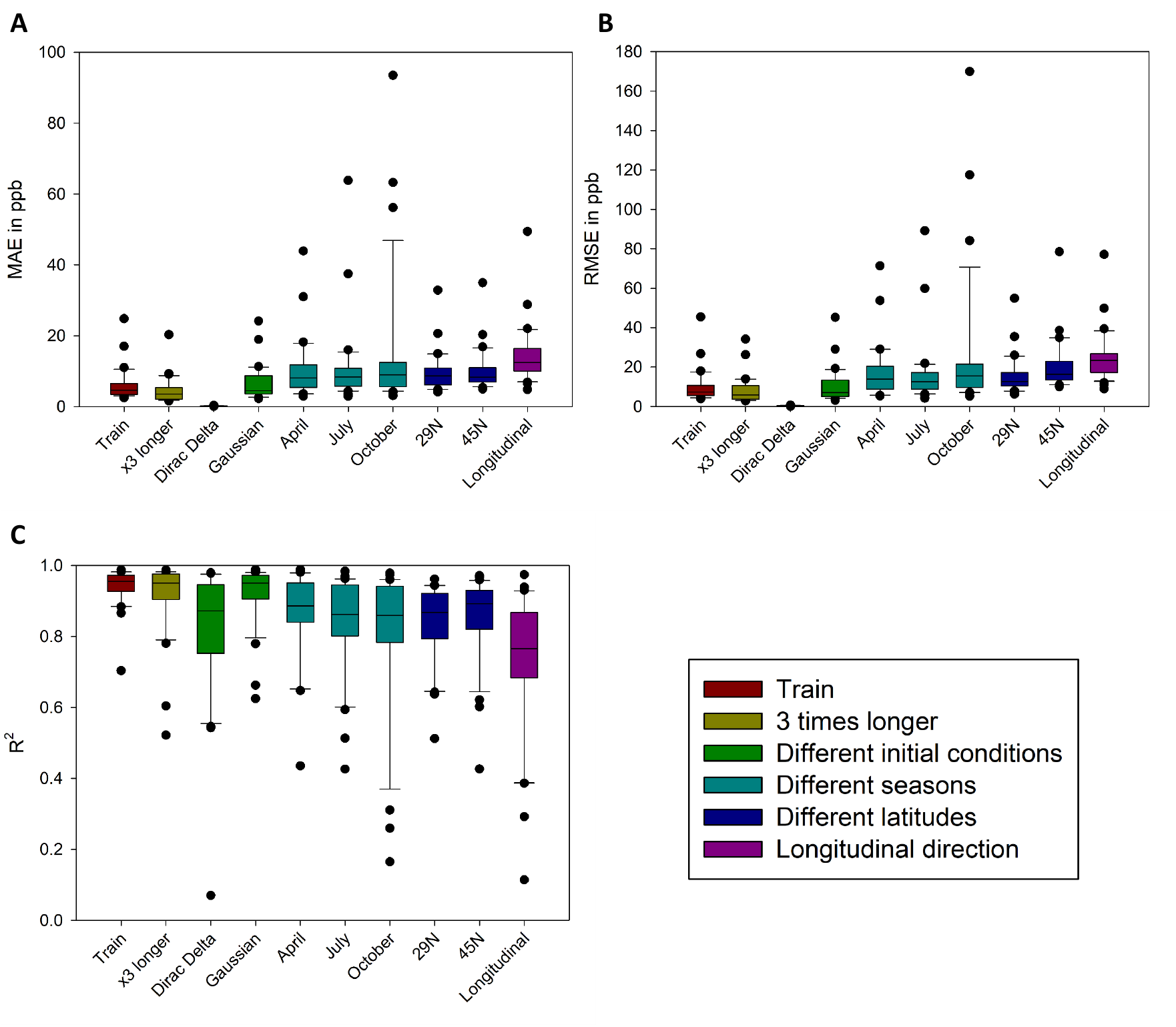}
    \caption{Box plots of error statistics in generalization tests for (A) MAE, (B) RMSE, and (C) R$^2$. Each data point represents one combination of spatial and temporal resolutions. Unstable simulations are not shown here but are shown in Figures~\ref{fig:Gen30}–\ref{fig:GenLong}.}
    \label{fig:GenSummary}
\end{figure}

Figure~\ref{fig:GenSummary} summarizes error statistics for the generalization tests described in Section~\ref{sec:generalization}. 
(Detailed results for each test are shown in Figures~\ref{fig:Gen30}\textnormal{,}~\ref{fig:GenIC}\textnormal{,}~\ref{fig:GenLat}\textnormal{,}~\ref{fig:GenSeason}\textnormal{,}~\textnormal{and}~\ref{fig:GenLong}). 
The numerical errors in generalization testing datasets are similar to the error in the training dataset in most cases, but outliers exist in the generalization tests with errors substantially larger than that found with the training dataset, and numerical instability does occur in some cases. 
In most of the unstable cases, we could resolve the instability by using model parameters from a training epoch different than the one that is optimal for the training dataset; those cases are shown in bold letters in Figures~\ref{fig:Gen30}\textnormal{,}~\ref{fig:GenIC}\textnormal{,}~\ref{fig:GenLat}\textnormal{,}~\ref{fig:GenSeason}\textnormal{,}~\textnormal{and}~\ref{fig:GenLong}. 
The goal of the analysis here is to explore the performance and flexibility of our method when trained on a small dataset, but we would expect to see better generalization if we used a more diverse training dataset.

In Figure~\ref{fig:GenSummary}, the ``3~times longer'' case shows that predictive performance does not significantly degrade over an extended prediction horizon. 
It also suggests that the learned solver would work well in integrating wind fields with similar characteristics (e.g., season and latitude) to the training dataset. 

The Dirac-delta initial condition test shows that changing the initial condition deteriorates the performance when the new initial condition is extraordinarily stiff (Figure~\ref{fig:GenSummary}). 
The normalized error is small, but this is because the initial mass is small, rather than indicating good model performance. 
The $R^{2}$ value better captures the large uncertainty of the learned emulation in this case. 
In contrast, the learned solver is robust to the initial condition with a smooth gradient (Gaussian shape; Figure~\ref{fig:GenSummary}). 
\citet{zhuang2021learned} found their learned 1-D advection scheme had a tendency to have a square-shaped scalar even though they fed the Gaussian distribution-like initial condition because their scheme was trained with the square wave moving to one direction. 
In our study, that does not happen because our solver could learn from not only the initial shape but also different spatial patterns with time evolution. 
However, our results still imply that training our model with a wider range of initial conditions would be useful. 

The effects of seasonality and different spatial domains are larger than other testing scenarios in terms of median error (Figure~\ref{fig:GenSummary}). 
The largest discrepancy in median appears in changing the spatial domain to a longitudinal direction. 
The largest outlier is seemingly shown in the October wind case, but this is only because the Figure~\ref{fig:GenSummary} does not show unstable simulations in the July and $\SI{29.5}{\degree}$N tests for (2$\Delta$x, 64$\Delta$t) resolution. 
As above, we expect that these issues could be resolved by training our model on a more diverse dataset that has different ranges of velocity, stiffness of gradient, etc. 

\begin{figure}[htp]
    \centering
    \includegraphics[width=0.5\textwidth]{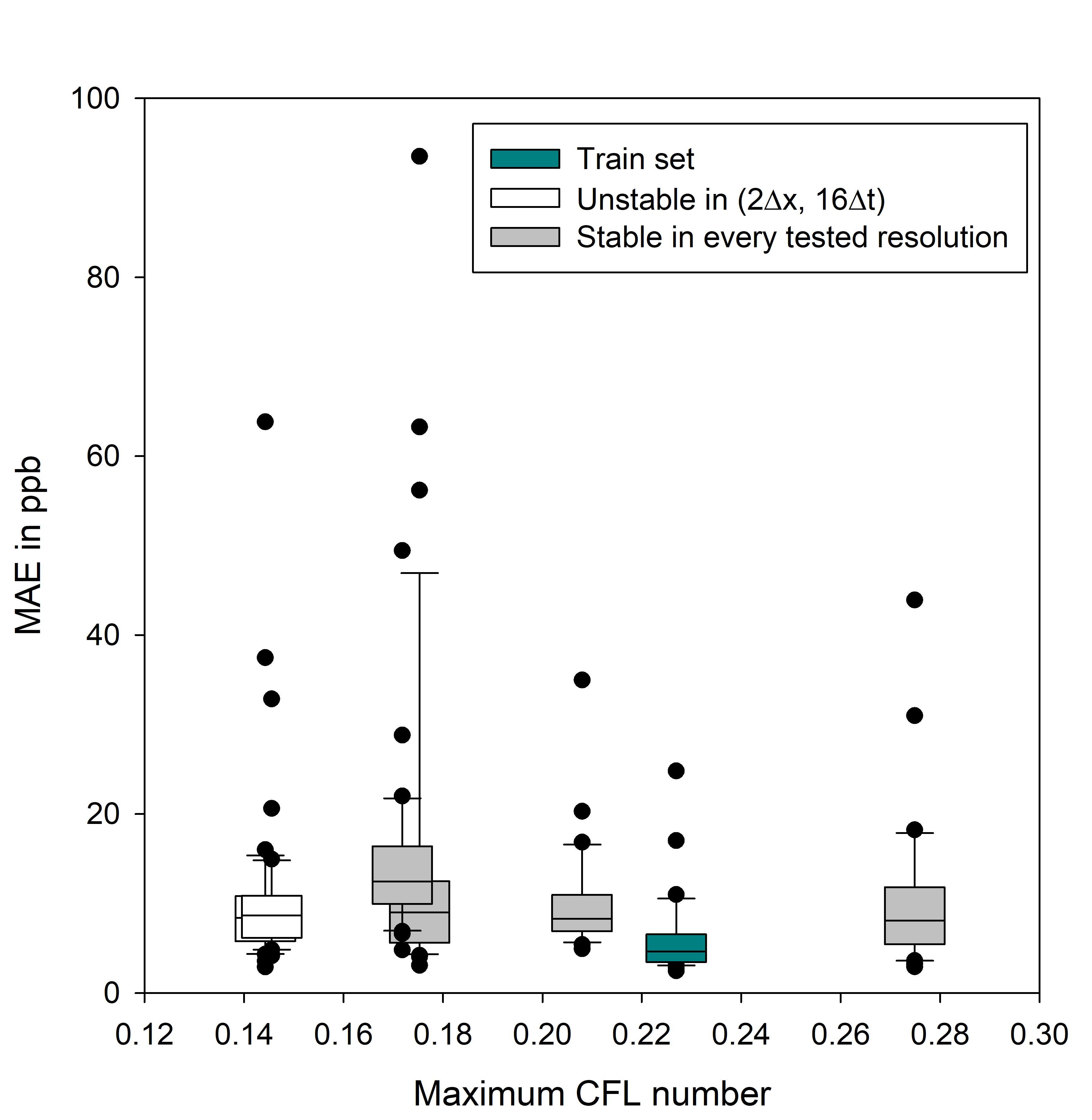}
    \caption{Dependence of generalization test accuracy on the maximum CFL number of target wind fields. Each box plot is the summary of errors in each testing scenario integrating wind data different from the training dataset.}
    \label{fig:CFL-MAE}
\end{figure}

To look closer into the effect of wind velocity range on the generalization skill of learned solvers, we plot MAE vs. maximum CFL number of the velocity fields used in the generalization testing (Figure~\ref{fig:CFL-MAE}). 
This plot shows that the learned solver tended to be unstable when the maximum CFL is far from the maximum CFL of the training dataset. 
The testing cases with the largest discrepancy in maximum CFL with the training dataset show numerical instabilities in (2$\Delta$x, 64$\Delta$t) resolution. 
This feature might originate from the structure of our learned solver. 
As shown in Figure~\ref{fig:ModelStructure}, the learned solver could scale the magnitude of scalar values but not in the case of the velocity field. 
Therefore, the neural network could fail to give the learned coefficients in a reasonable range when the input velocity is not close to the training regime. 
Based on this intuition, in future work, we will want to train the solver with a larger dataset that has a wider range of velocity. 
However, as mentioned above, our goal here is to explore the potential of this approach rather than producing a highly-tuned 1-D advection solver. 

Overall, the results from our 1-D generalization tests suggest that our learned scheme can work in a relatively wide set of circumstances, even when trained on an extremely limited dataset. 
Our results additionally suggest that it is important for the training dataset to include a distribution of CFL numbers similar to the conditions the model is expected to be deployed under. 
This will likely be even more crucial in 3-D applications where wind speed varies greatly with altitude. 

\subsection{2-D demonstration}\label{sec:multidimesionresult}
\begin{figure}[htp]
    \centering
    \includegraphics[width=\textwidth]{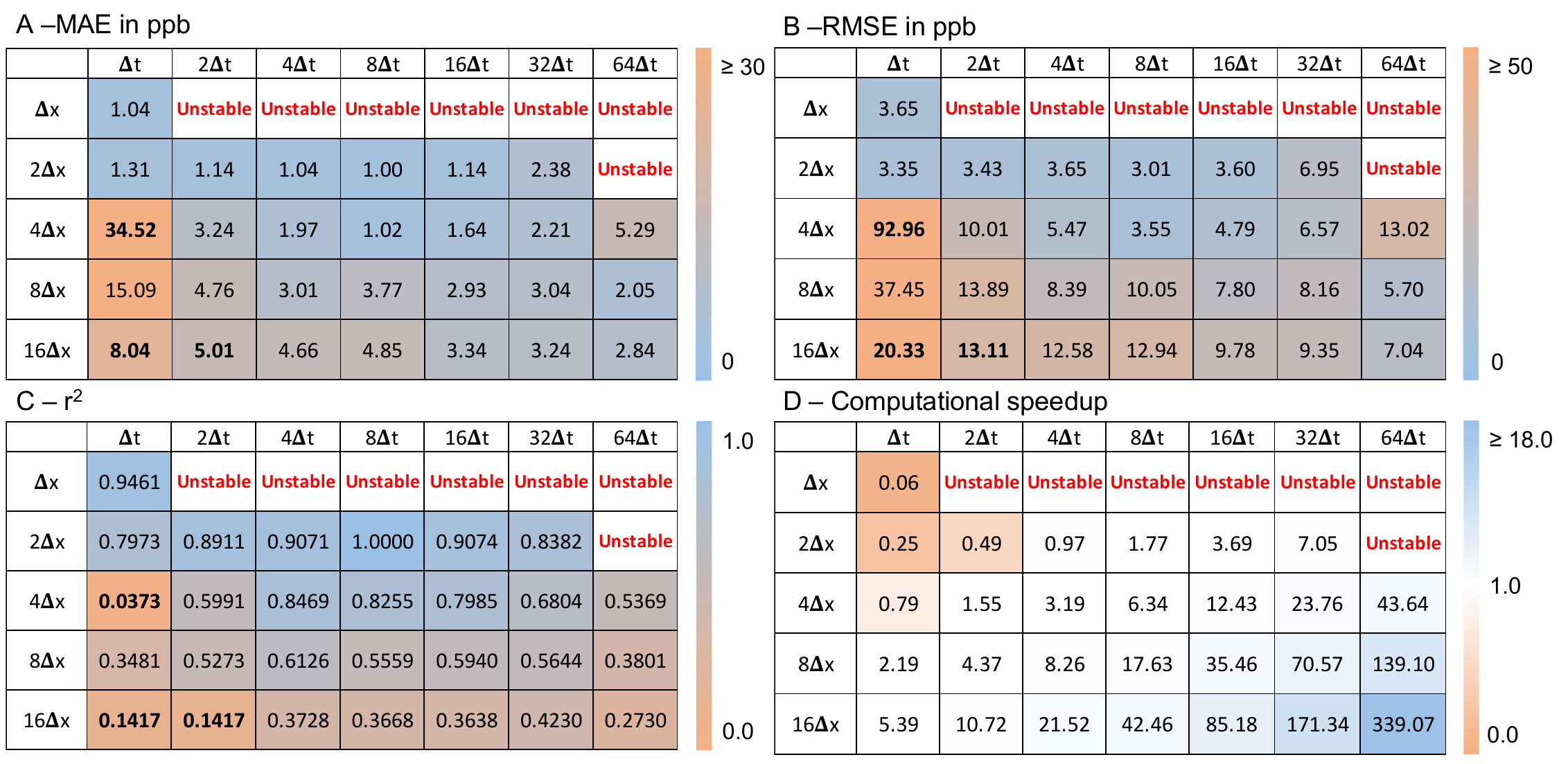}
    \caption{Performance of the learned solver in 2-D implementation in different coarse-graining resolutions. Cases with bold numbers were unstable with the optimal training epoch but stable when using a different training epoch.}
    \label{fig:2DLearnedStats}
\end{figure}

Figure~\ref{fig:2DLearnedStats} shows the performance of the learned 1-D advection solvers when implemented in a 2-D simulation. 
Even though our learned solvers are trained with 1-D, ten among the 35 2-D simulation cases achieve R$^{2}$ values above 0.8. 
This is a fascinating result given that the velocity fields used in 2-D advection have more complex and diverse features than exist in our 1-D training dataset. 
Simply, the 2-D advection requires 224 $\times$ 201 times of 1-D advection, and the relative variety of wind patterns in the 2-D dataset could be a function of that dimension. 
In this regard, our approach looks promising even in 2-D applications. 
The best accuracy is observed in (2$\Delta x$, 8$\Delta t$), which shows remarkable performance including R$^{2}$ = 1.00. 
The maximal acceleration is 340~times in the coarsest resolution. 
Computational time statistics for 2-D advection by the learned solvers are given in Table~\ref{table:time_learned_2D}.

Although the fact that our 1-D solver works at all in a 2-D context has exceeded our expectations, we nonetheless proceed with an analysis of strengths and weaknesses with the goal of informing future work. 
Unlike the results in 1-D (Figure~\ref{fig:TrainStats}), 2-D simulations in several resolutions show poor results (R$^{2} < 0.5$). 
There are unstable simulations in 2-D as well:
 simulations in (1$\Delta x$, 2$\Delta t$), (1$\Delta x$, 4$\Delta t$), (1$\Delta x$, 32$\Delta t$), and (2$\Delta x$, 64$\Delta t$) are unstable in 2-D advection, though they are stable in 1-D. 
We suspect that this limited performance may be caused by the existence of conditions (for example a greater range of wind speeds) that were not present in the training dataset.

\begin{figure}[htp]
    \centering
    \includegraphics[width=\textwidth]{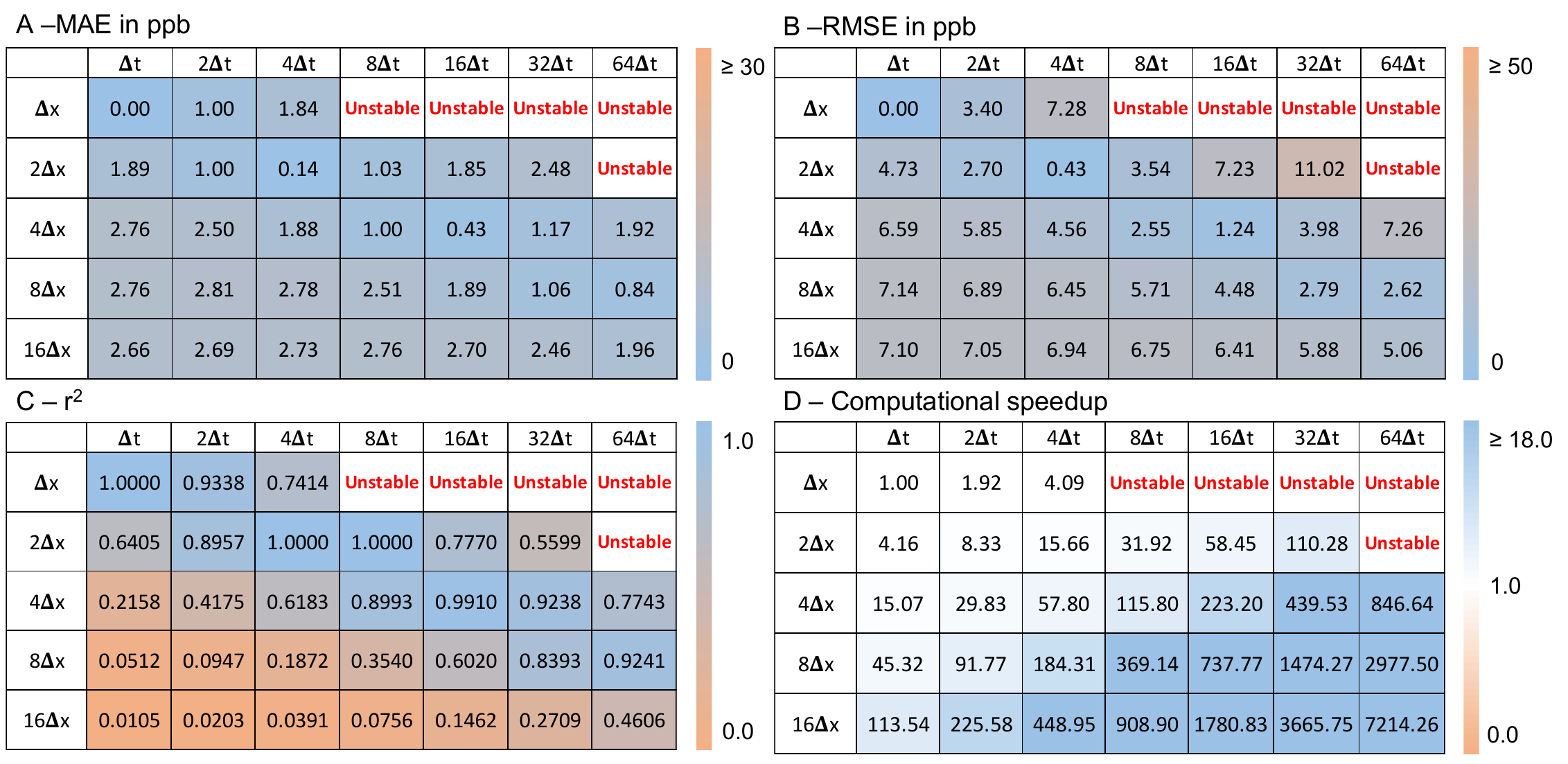}
    \caption{Performance of the reference solver in 2-D at different resolutions.}
    \label{fig:2DNumericalStats}
\end{figure}

To fairly evaluate our approach, we also test the 2-D simulations with the low-resolution reference solver. 
Similar to 1-D coarsening cases, 2-D advection using the reference solver shows good performance in the diagonal elements of the coarsening performance matrix (Figure~\ref{fig:2DNumericalStats}). 
In other coarsening regimes, the reference solvers show a poor R$^{2}$ but usually maintain small errors. 
This is because the reference solver has the local slope limiter to satisfy the total variance diminishing property, thus the simulation is free from spurious oscillation when the simulation does not violate the CFL condition \citep{durran2010numerical}. 
Instead, the solution dissipates to zero due to numerical diffusion, resulting in low error but also low correlation. 
Instability occurs when the coarsening results in the CFL number greater than one. 
Numerical techniques exist that can handle conditions with CFL $> 1$, but they lie outside of the focus of the current study. 
The low-resolution reference solvers also achieve very good performance in some resolutions including (2$\Delta x$, 4$\Delta t$) and (2$\Delta x$, 8$\Delta t$). 
The computational speedup by coarsening is always greater with the reference solver than with the learned solver in the same resolution, as explained in Section~\ref{sec:trainresult}. 
The maximum acceleration in 2-D by the reference solver is $7200\times$. 
Table~\ref{table:time_reference_2D} summarizes the computational time for 2-D advection by the reference solvers in different scales.

\begin{figure}[htp]
    \centering
    \includegraphics[width=\textwidth]{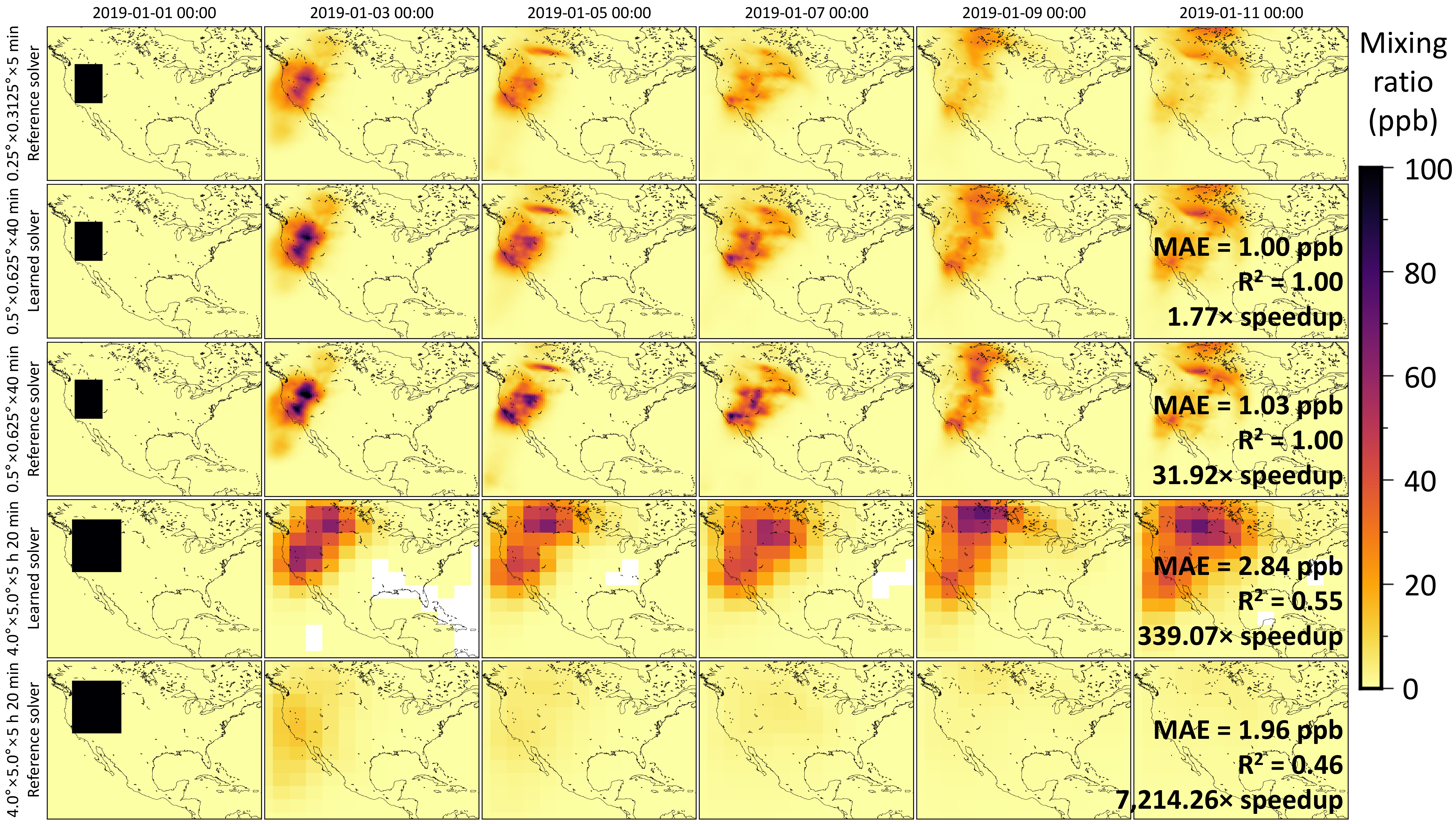}
    \caption{2-D demonstration of coarse-graining by the learned and reference advection schemes. The first row shows the results of the high-resolution baseline solver. The second and fourth rows are the learned advection in different resolutions. The third and fifth rows are the low-resolution reference solvers at different resolutions.}
    \label{fig:2D}
\end{figure}

We visualize the 2-D implementation results of both reference and learned solvers at two resolutions (Figure~\ref{fig:2D}). 
We select (2$\Delta x$, 8$\Delta t$) and (16$\Delta x$, 64$\Delta t$), representing the best performance and the maximum acceleration by the learned solvers respectively. 
In (2$\Delta x$, 8$\Delta t$), both solvers almost perfectly emulate the finest grid simulation. 
Both solvers have a tendency to exaggerate the thickness of the plume, but overall they represent the spatial pattern well. 
For (16$\Delta x$, 64$\Delta t$), we additionally feed the 1-D generalization testing datasets to the model as well as the original training dataset to enhance the learned model. 
Originally, the learned scheme trained only with the training dataset have limited performance with R$^{2}$ = 0.27 (Figure~\ref{fig:2DLearnedStats}). 
With the increased training data, the learned scheme R$^{2}$ is doubled, showing the benefit of introducing more data into the learned model. 
However, the learned scheme in this resolution overestimates the concentration of the plume and sometimes shows a negative concentration (with a magnitude less than $1$~ppb). 
These artifacts result in a relatively large MAE value. 
The reference solver in this resolution showed a similar range of R$^{2}$, but dissipates the plume too quickly and the overall concentration range near the plume is much too low. 
However, even though the reference solver is dissipative at this resolution, the MAE value was still smaller than the learned solver. 

\section{Discussion}\label{sec:discussion}
We will discuss the approach described herein in terms of strengths, weaknesses, and directions for future research. 
The first noticeable strength is the decent performance of the learned model in emulating the reference model, even without using a large training dataset or hyperparameter tuning. 
The coarse-grained learned solver achieves accuracy that is similar to or better than the corresponding-resolution reference solver for conditions similar to the training dataset and performs well in some cases where the reference solver does not perform well, particularly for simulations with temporal coarsening but without corresponding spatial coarsening. 
This is a key benefit of our approach because for many applications of air quality modeling, high spatial resolution is much more important than high temporal resolution. 
For example, CTMs typically save results to disk every one or three simulation hours, discarding all information at a higher temporal resolution than that. 
Therefore, an approach that allows temporal coarse-graining without corresponding spatial coarse-graining could achieve computational acceleration without requiring any reduction in the detail of the CTM output.

Furthermore, our learned solver generalizes fairly well to conditions it was not trained on, even though it was trained on an extremely limited amount of data. 
This good generalization performance extends to the application of our learned model for 2-D simulations, even though it was only trained in 1-D. 
Finally, our learned solvers can achieve substantial computational acceleration relative to the reference solver, with maximum acceleration at $\sim10\times$ for 1-D and $\sim100\times$ for 2-D simulations. 
Together, these results suggest that a scaled-up version of the approach shown here could be a good candidate for embedding within a full CTM for further testing, and this is a priority area for future work.

One limitation of our learned advection model is that it is numerically unstable under certain conditions. 
The number of scenarios where numerical instability occurred was similar for the learned and reference solvers, but the conditions that cause instability in the learned solver are somewhat unpredictable, whereas with the reference solver instability predictably occurs at conditions with the highest CFL values. 

This limitation could be addressed in future work by using a larger and more diverse training dataset to reduce the chance of the solver encountering conditions that were not well represented during training, or by minimizing error over a larger number of timesteps to encourage stability over longer time horizons. 
Fine-tuning the hyperparameters could also help to develop a more accurate and stable model. 
Other possibilities could be to bound in the magnitude of the learned coefficients, similar to the idea of a flux limiter, or to add terms to the training loss function to encourage mass conservation or stability (e.g., the amplification ratio in von Neumann analysis should be less than or equal to one).

We consider a learned solver to be useful when it is faster than the high-resolution reference solver and more accurate than the equivalent-resolution reference solver. 
A second limitation of the current study is that this is only true for conditions where the reference solver is particularly inaccurate, i.e. simulations with large timesteps and small grid cells or simulations with small grid cells and large timesteps (Figure~\ref{fig:NumericalStats}).

It may not ever be reasonable to expect a learned solver to be faster than the reference solver when run at the same resolution because in general, a learned model will have more degrees of freedom than a theoretically derived model, and the additional degrees of freedom require additional computational operations. 
However, future work could reduce the computational intensity of the learned model by exploring different machine-learning architectures for predicting the operator coefficients ($a$ and $b$ in Equation \ref{eq:learnedsolver}). 
A learned model with a smaller per-step computational penalty as compared to the reference solver would have a larger range of conditions where it was useful.

\section{Conclusions}\label{sec:conclusion}

In this study, we train a learned 1-D advection solver with 10 days of transport data and characterize its performance with the goal of exploring its general strengths and weaknesses rather than optimizing its performance for a particular use case. 
Even with this limited training dataset, the learned 1-D advection scheme shows generally stable and accurate emulation which is robust against unforeseen conditions. 
The 1-D learned solver could run the simulation up to 18 times faster than the baseline solver and could be used in the 2-D advection with an appropriate splitting technique. 
The 2-D implementation results are still mostly stable, reasonably accurate, and overall more successful than we expected ahead of time. 
Our approach demonstrates up to 340$\times$ computational speedup in the 2-D application. 
For simulations with temporal coarse-graining but without spatial coarse-graining, numerical instability occurred occasionally. 
Addressing this instability in future research will help to fully realize the potential of this approach. 
We hope that the findings in this study serve as an inspiration and a benchmark for future studies on multidimensional learned advection operators.

%

%

\clearpage
\acknowledgments
This work is supported by an Early Career Faculty grant from the National Aeronautics and Space Administration (grant no. 80NSSC21K1813) and Assistance Agreement RD-84001201-0  awarded by the U.S. Environmental Protection Agency.  It has not been formally reviewed by EPA. The views expressed in this document are solely those of authors and do not necessarily reflect those of the Agency.  EPA does not endorse any products or commercial services mentioned in this publication. MP is supported by the Carver Fellowship and Illinois Distinguished Fellowship.

%
%
\datastatement
The code to implement the reference and learned advection and the velocity dataset are available at \url{https://doi.org/10.5281/zenodo.8111623}. 
The output of the model training including the model parameters and outputs will be available at \url{https://doi.org/10.13012/B2IDB-4743181_V1} and currently can be accessed via \url{https://databank.illinois.edu/datasets/IDB-4743181?code=8zCTeEEae1cO3_3JNLKndtCk4h9FwhzG2L6iCqhGzDU}.

\clearpage







%



\appendix 
\appendixtitle{Appendix Tables and Figures}
\subsection{Appendix Tables}
\begin{table}[h!]
\caption{The best training epoch in each resolution tested in this study. In table cells with bold-face numbers, the training was not successful with 3-layer CNN but was successful with 4-layer CNN. Cases marked ``unstable'' were unable to perform stable simulations even with a 5-layer CNN.}\label{table:TrainEpoch}
\begin{center}
\begin{tabular}{cccccccc}
\topline
& $\Delta t$ & 2$\Delta t$ & 4$\Delta t$ & 8$\Delta t$ & 16$\Delta t$ & 32$\Delta t$ & 64$\Delta t$\\
\midline
 $\Delta x$ & 95 & 112 & \textbf{13} & Unstable & Unstable & \textbf{225} & Unstable\\
 2$\Delta x$ & 71 & 59 & 31 & 164 & 137 & 282 & 221\\
 4$\Delta x$ & 57 & 33 & 75 & 197 & 100 & 199 & 100\\
 8$\Delta x$ & 193 & 443 & 362 & 23 & 43 & 44 & 263\\
 16$\Delta x$ & 65 & 166 & 85 & 31 & 41 & 25 & 56\\
\botline
\end{tabular}
\end{center}
\end{table}
\clearpage

\begin{table}[h!]
\caption{Computational time taken to integrate 10-day-long 1-D advection by the learned advection solvers. (Unit: milliseconds; *: measured only once because the computational time is large enough; N/A: not measured because the solver is unstable)}\label{table:time_learned}
\resizebox{\columnwidth}{!}{
\begin{tabular}{cccccccc}
\topline
& $\Delta t$ & 2$\Delta t$ & 4$\Delta t$ & 8$\Delta t$ & 16$\Delta t$ & 32$\Delta t$ & 64$\Delta t$\\
\midline
 $\Delta x$ & 720.9 $\pm$ 9.8 & 365.5 $\pm$ 5.6 & 15722$^*$ & N/A & N/A & 1699 $\pm$ 83.3 & N/A\\
 2$\Delta x$ & 407.3 $\pm$ 7.4 & 207.4 $\pm$ 8.6 & 104.6 $\pm$ 2.8 & 50.9 $\pm$ 2.3 & 27.7 $\pm$ 2.2 & 12.3 $\pm$ 1.4 & 6.6 $\pm$ 1.1\\
 4$\Delta x$ & 252.2 $\pm$ 7.1 & 125.8 $\pm$ 2.8 & 62.5 $\pm$ 3.0 & 31.9 $\pm$ 2.9 & 15.9 $\pm$ 2.3 & 8.4 $\pm$ 1.6 & 3.8 $\pm$ 1.1\\
 8$\Delta x$ & 173.0 $\pm$ 6.2 & 86.6 $\pm$ 4.2 & 43.1 $\pm$ 3.9 & 21.5 $\pm$ 2.9 & 10.7 $\pm$ 2.1 & 5.4 $\pm$ 1.5 & 2.8 $\pm$ 1.0\\
 16$\Delta x$ & 134.6 $\pm$ 5.4 & 68.4 $\pm$ 7.7 & 33.7 $\pm$ 5.3 & 16.7 $\pm$ 3.8 & 8.3 $\pm$ 2.5 & 4.1 $\pm$ 1.7 & 2.1 $\pm$ 1.2\\
\botline
\end{tabular}}
\end{table}
\clearpage

\begin{table}[h!]
\caption{Computational time taken to integrate 10-day-long 1-D advection by the learned advection solvers.  (Unit: milliseconds; *: measured only once because the computational time is large enough; N/A: not measured because the solver is unstable)}\label{table:time_reference}
\begin{center}
\begin{tabular}{cccccccc}
\topline
 & $\Delta t$ & 2$\Delta t$ & 4$\Delta t$ & 8$\Delta t$ & 16$\Delta t$ & 32$\Delta t$ & 64$\Delta t$\\
\midline
 $\Delta x$ & 37.9 $\pm$ 7.4 & 19.5 $\pm$ 6.1 & 9.5 $\pm$ 3.9 & 4.8 $\pm$ 2.8 & 2.4 $\pm$ 1.9 & 1.2 $\pm$ 1.4 & 0.6 $\pm$ 1.0\\
 2$\Delta x$ & 19.9 $\pm$ 5.4 & 9.9 $\pm$ 3.9 & 5.0 $\pm$ 2.8 & 2.5 $\pm$ 2.0 & 1.2 $\pm$ 1.4 & 0.6 $\pm$ 1.0 & 0.3 $\pm$ 0.7\\
 4$\Delta x$ & 11.2 $\pm$ 4.4 & 5.6 $\pm$ 3.1 & 2.8 $\pm$ 2.1 & 1.4 $\pm$ 1.6 & 0.7 $\pm$ 1.1 & 0.4 $\pm$ 0.8 & 0.2 $\pm$ 0.5\\
 8$\Delta x$ & 7.0 $\pm$ 3.4 & 3.6 $\pm$ 2.6 & 1.8 $\pm$ 1.7 & 0.9 $\pm$ 1.2 & 0.4 $\pm$ 0.9 & 0.2 $\pm$ 0.6 & 0.1 $\pm$ 0.4\\
 16$\Delta x$ & 4.8 $\pm$ 3.1 & 2.4 $\pm$ 2.1 & 1.2 $\pm$ 1.6 & 0.6 $\pm$ 1.0 & 0.3 $\pm$ 0.7 & 0.2 $\pm$ 0.5 & 0.1 $\pm$ 0.3\\
\botline
\end{tabular}
\end{center}
\end{table}
\clearpage

\begin{table}[h!]
\caption{The maximum CFL number of each wind field tested in this study.}\label{table:MaxCFL}
\begin{center}
\begin{tabular}{cccccccc}
\topline
 & $\Delta t$ & 2$\Delta t$ & 4$\Delta t$ & 8$\Delta t$ & 16$\Delta t$ & 32$\Delta t$ & 64$\Delta t$\\
\midline
 $\Delta x$ & 0.23 & 0.45 & 0.91 & 1.82 & 3.63 & 7.26 & 14.52\\
 2$\Delta x$ & 0.11 & 0.23 & 0.45 & 0.91 & 1.81 & 3.63 & 7.26\\
 4$\Delta x$ & 0.06 & 0.11 & 0.23 & 0.45 & 0.90 & 1.81 & 3.61\\
 8$\Delta x$ & 0.03 & 0.06 & 0.11 & 0.22 & 0.45 & 0.90 & 1.79\\
 16$\Delta x$ & 0.01 & 0.03 & 0.05 & 0.11 & 0.22 & 0.43 & 0.87\\
\botline
\end{tabular}
\end{center}
\end{table}
\clearpage

\begin{table}[h!]
\caption{Computational time taken to integrate 10-day-long 2-D advection by the learned advection solvers. (Unit: milliseconds; *: measured only once because the computational time is large enough; N/A: not measured because the solver is unstable)}\label{table:time_learned_2D}
\resizebox{\columnwidth}{!}{
\begin{tabular}{cccccccc}
\topline
 & $\Delta t$ & 2$\Delta t$ & 4$\Delta t$ & 8$\Delta t$ & 16$\Delta t$ & 32$\Delta t$ & 64$\Delta t$\\
\midline
 $\Delta x$ & 593,080$^*$ & 289,616$^*$ & 9,362,782$^*$ & N/A & N/A & N/A & N/A\\
 2$\Delta x$ & 152,659$^*$ & 76,292$^*$ & 38,424$^*$ & 21,179$^*$ & 10,158$^*$ & 5,311$^*$ & N/A\\
 4$\Delta x$ & 47,217$^*$ & 24,102$^*$ & 11,722$^*$ & 5,903$^*$ & 3,013 $\pm$ 59.7 & 1,576 $\pm$ 14.7 & 857.9 $\pm$ 6.8\\
 8$\Delta x$ & 17,075$^*$ & 8,559$^*$ & 4,531 $\pm$ 405.4 & 2,124 $\pm$ 26.2 & 1,056 $\pm$ 12.9 & 530.5 $\pm$ 7.3 & 269.2 $\pm$ 7.8\\
 16$\Delta x$ & 6,942$^*$ & 3,494 $\pm$ 31.3 & 1,740 $\pm$ 9.8 & 881.8 $\pm$ 11.2 & 439.6 $\pm$ 12.3 & 218.5 $\pm$ 8.5 & 110.4 $\pm$ 10.2\\
\botline
\end{tabular}}
\end{table}
\clearpage

\begin{table}[h!]
\caption{Computational time taken to integrate 10-day-long 2-D advection by the learned advection solvers. (Unit: milliseconds)}\label{table:time_reference_2D}
\resizebox{\columnwidth}{!}{
\begin{tabular}{cccccccc}
\topline
 & $\Delta t$ & 2$\Delta t$ & 4$\Delta t$ & 8$\Delta t$ & 16$\Delta t$ & 32$\Delta t$ & 64$\Delta t$\\
\midline
 $\Delta x$ & 37,442$^*$ & 19,531$^*$ & 9,162$^*$ & N/A & N/A & 1,323 $\pm$ 181.1 & N/A\\
 2$\Delta x$ & 8,997$^*$ & 4,497 $\pm$ 42.2 & 2,391 $\pm$ 220.8 & 1,173 $\pm$ 9.1 & 640.6 $\pm$ 94.0 & 339.5 $\pm$ 5.8 & N/A\\
 4$\Delta x$ & 2,485 $\pm$ 31.1 & 1,255 $\pm$ 7.5 & 647.8 $\pm$ 9.4 & 323.3 $\pm$ 9.8 & 167.8 $\pm$ 8.9 & 85.2 $\pm$ 2.1 & 44.2 $\pm$ 8.8\\
 8$\Delta x$ & 826.1 $\pm$ 11.6 & 408.0 $\pm$ 7.1 & 203.1 $\pm$ 8.0 & 101.4 $\pm$ 9.0 & 50.8 $\pm$ 2.6 & 25.4 $\pm$ 6.0 & 12.6 $\pm$ 3.0\\
 16$\Delta x$ & 329.8 $\pm$ 8.1 & 166.0 $\pm$ 5.8 & 83.4 $\pm$ 8.7 & 41.2 $\pm$ 7.7 & 21.0 $\pm$ 5.7 & 10.2 $\pm$ 3.6 & 5.2 $\pm$ 2.2\\
\botline
\end{tabular}}
\end{table}
\clearpage

\subsection{Appendix Figures}

\begin{figure}[htp]
    \centering
    \includegraphics[width=0.8\textwidth]{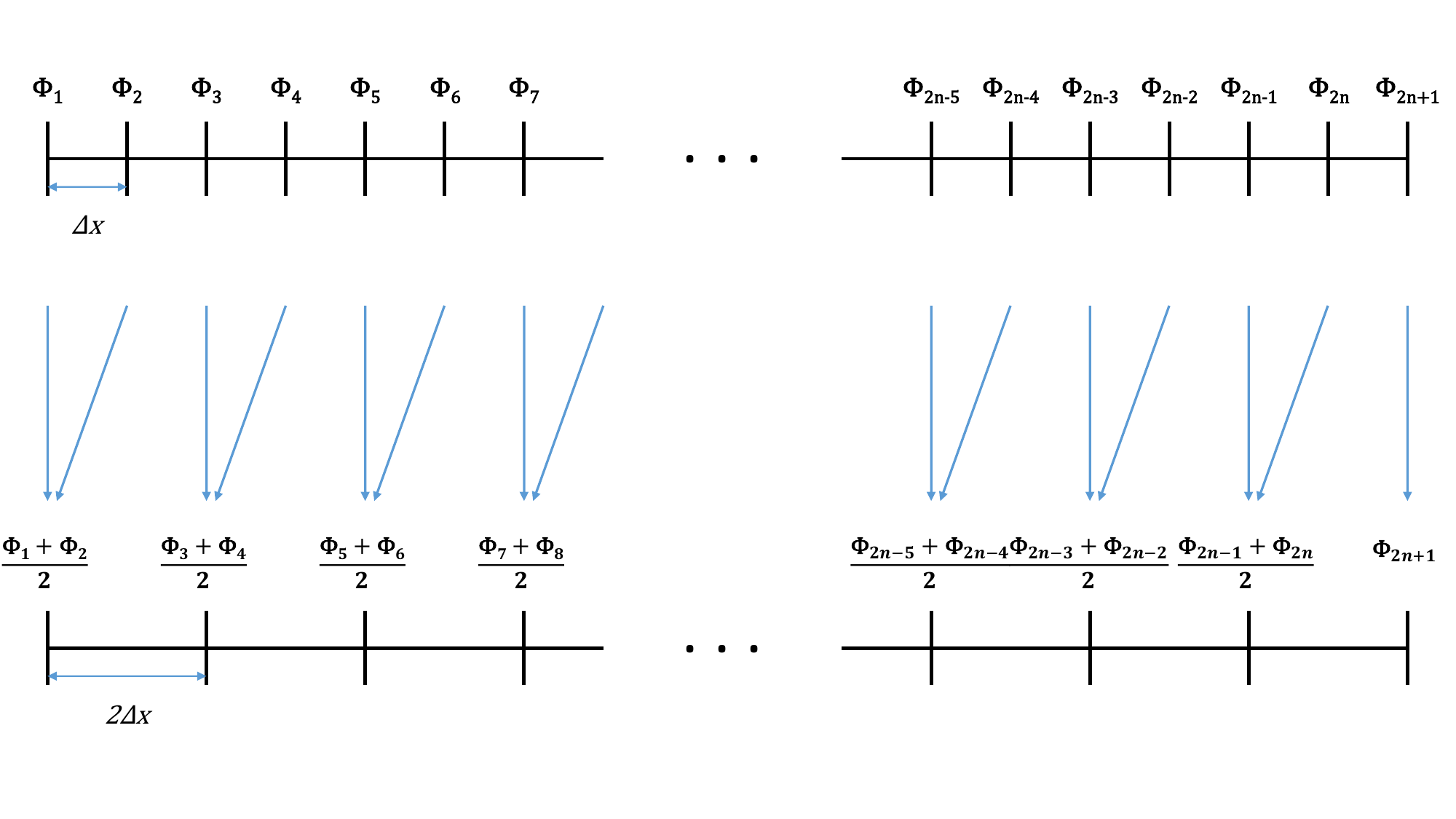}
    \caption{Illustration of the mass-conserving downsampling method.}
    \label{fig:Downsample}
\end{figure}
\clearpage

\begin{figure}[htp]
    \centering
    \includegraphics[width=\textwidth]{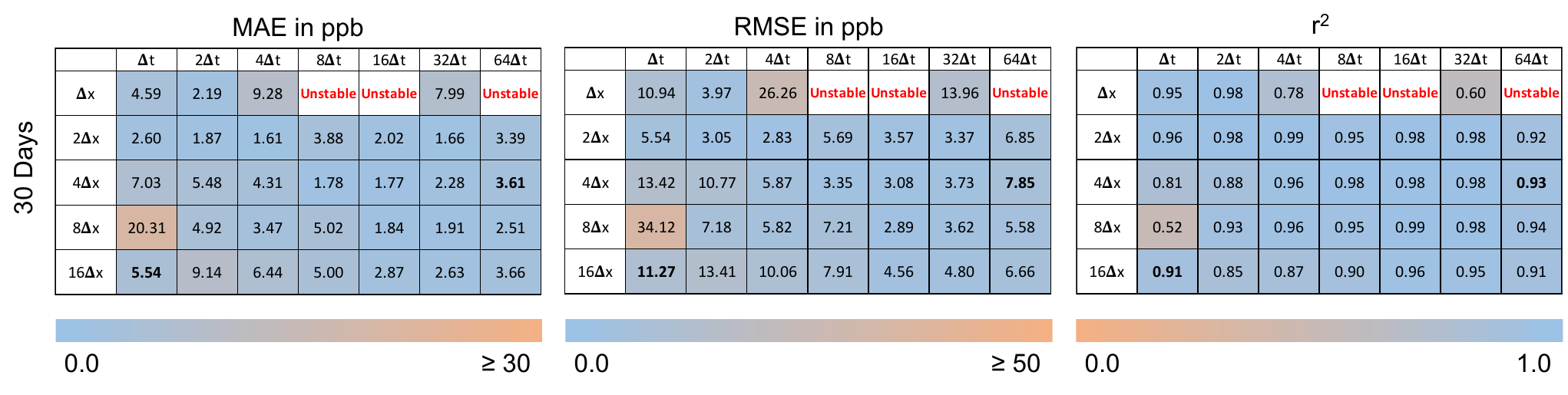}
    \caption{Summary of generalization testing with a longer time span. The cases with bold numbers were unstable when using the optimal epoch for the training data but stable when using a different epoch.}
    \label{fig:Gen30}
\end{figure}
\clearpage

\begin{figure}[htp]
    \centering
    \includegraphics[width=\textwidth]{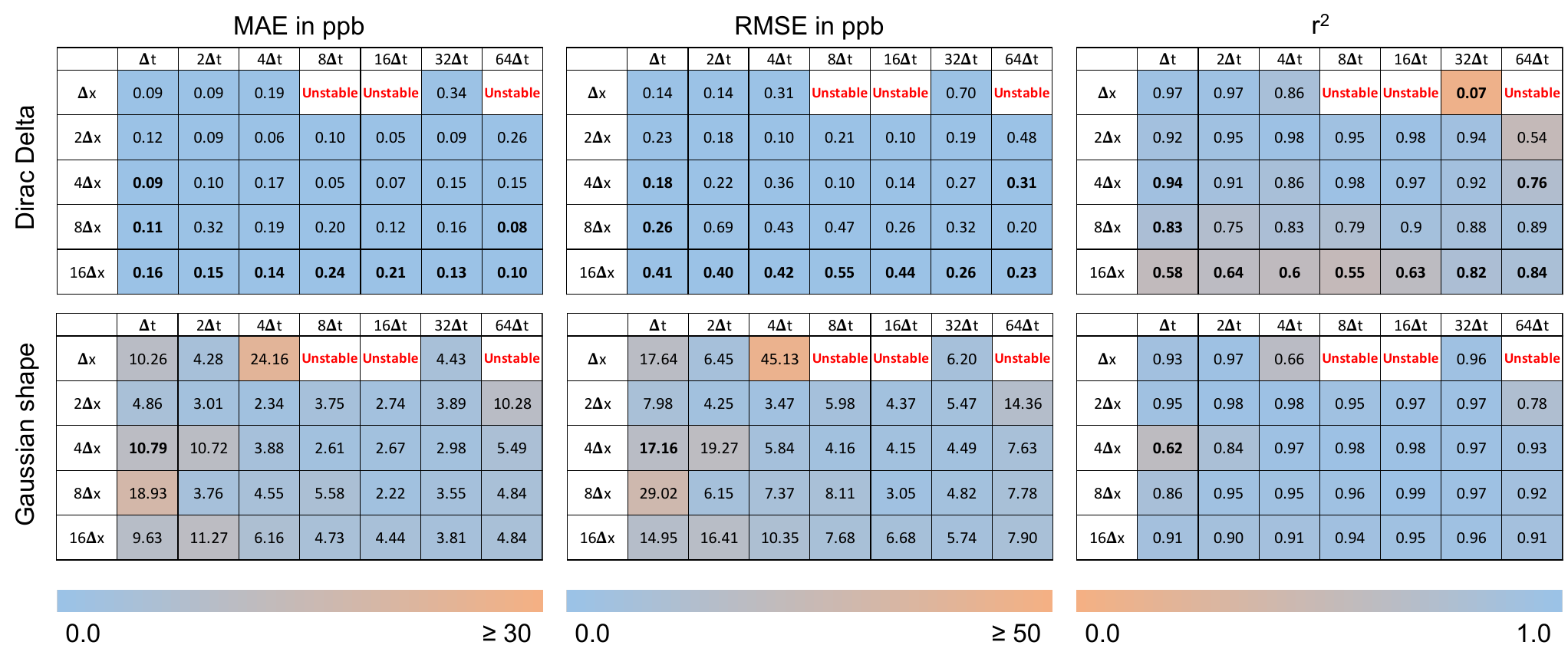}
    \caption{Summary of generalization testing with different initial conditions. The cases with the bold numbers were unstable when using the optimal epoch for the training data but stable when using a different training epoch.}
    \label{fig:GenIC}
\end{figure}
\clearpage

\begin{figure}[htp]
    \centering
    \includegraphics[width=\textwidth]{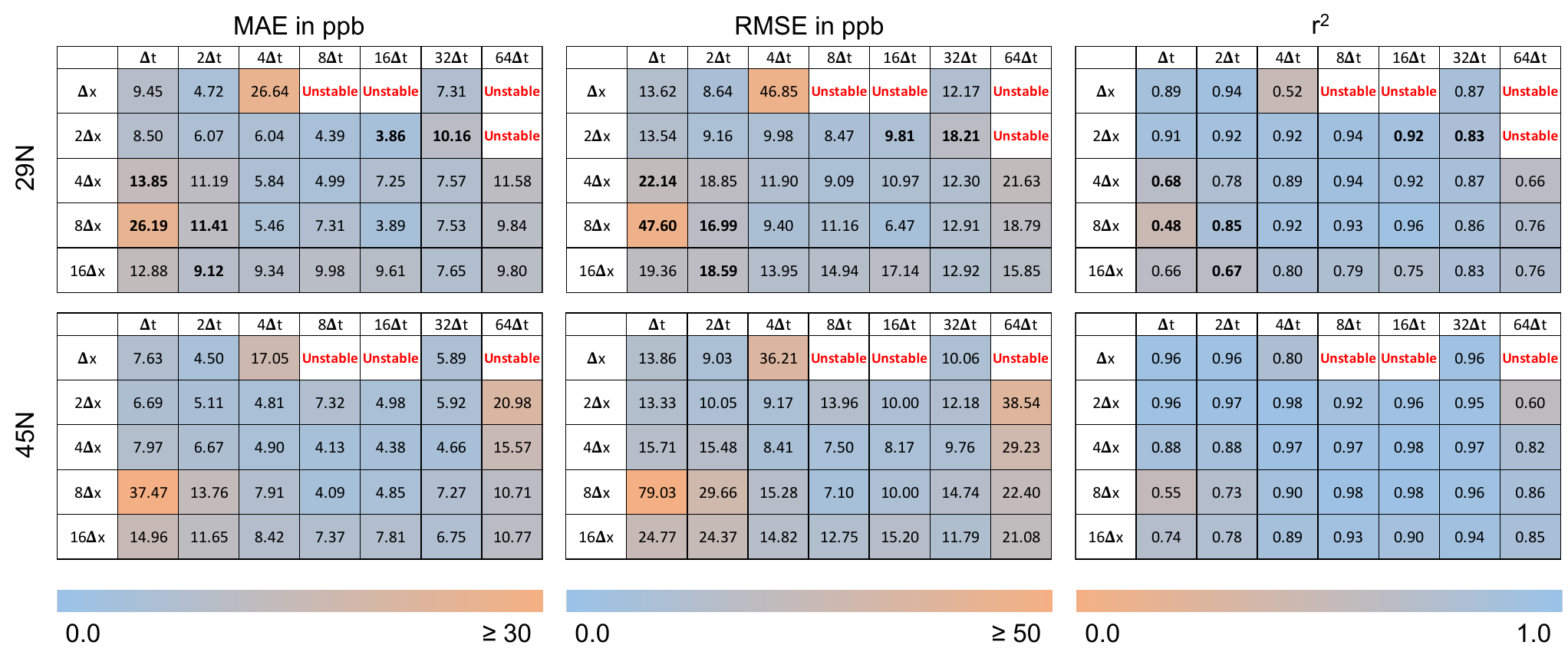}
    \caption{Summary of generalization testing with different latitudes. The cases with the bold numbers were unstable when using the optimal epoch for the training data but stable when using a different training epoch.}
    \label{fig:GenLat}
\end{figure}
\clearpage

\begin{figure}[htp]
    \centering
    \includegraphics[width=\textwidth]{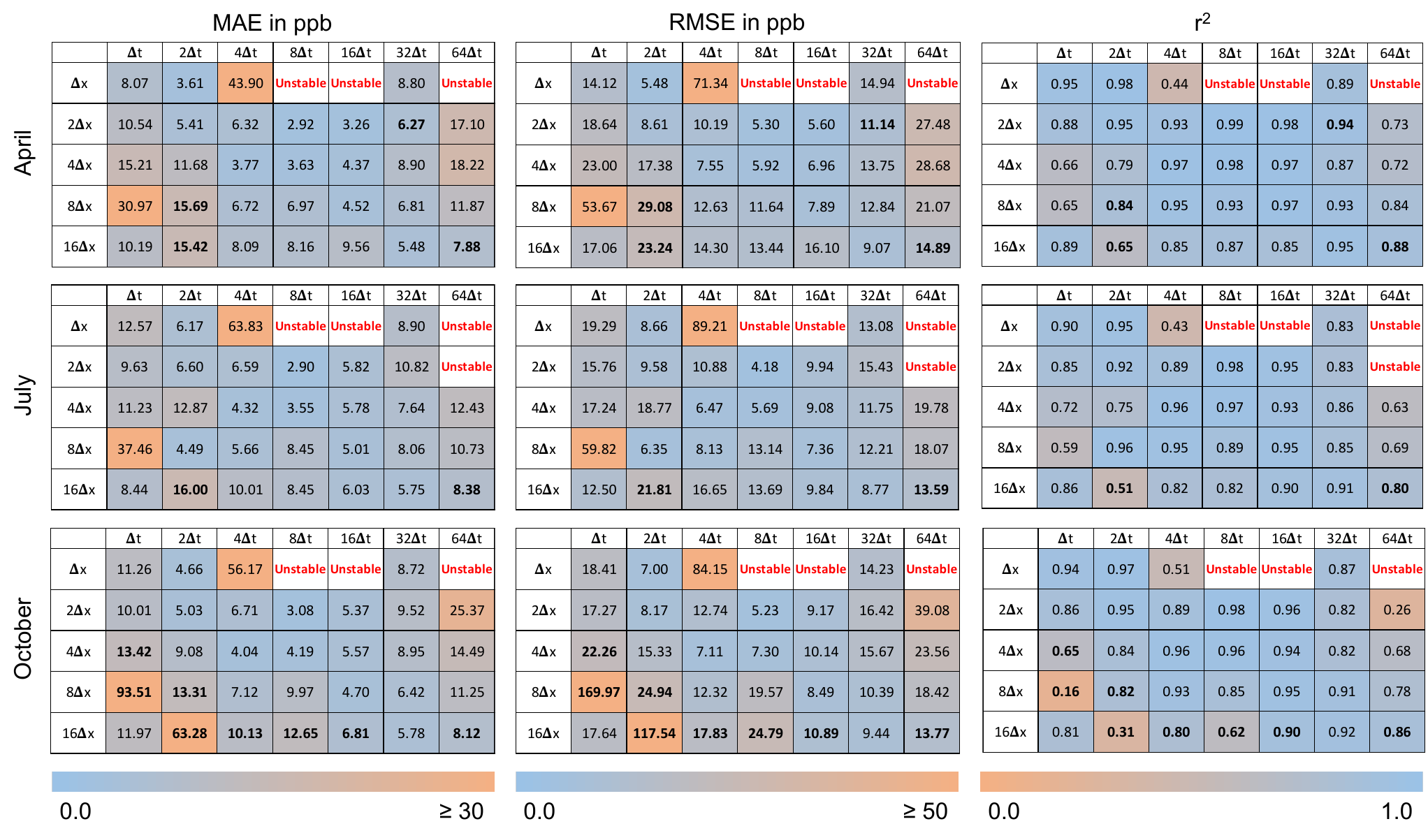}
    \caption{Summary of generalization testing with different seasons. The cases with the bold numbers were unstable when using the optimal epoch for the training data but stable when using a different training epoch.}
    \label{fig:GenSeason}
\end{figure}
\clearpage

\begin{figure}[htp]
    \centering
    \includegraphics[width=\textwidth]{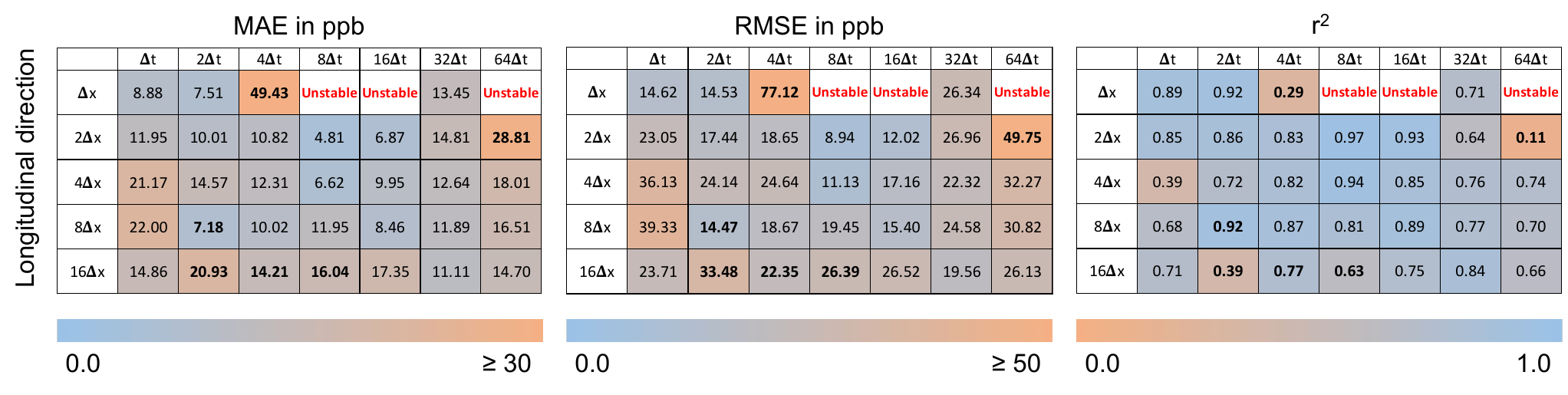}
    \caption{Summary of generalization testing with the longitudinal application. The cases with the bold numbers were unstable when using the optimal epoch for the training data but stable when using a different training epoch.}
    \label{fig:GenLong}
\end{figure}
\clearpage


\bibliographystyle{ametsocV6}
\bibliography{references}

\end{document}